\documentclass{SciPost}

\hypersetup{
    colorlinks,
    linkcolor={red!50!black},
    citecolor={blue},
    urlcolor={blue!80!black}
}

\DeclareSymbolFont{usualmathcal}{OMS}{cmsy}{m}{n}
\DeclareSymbolFontAlphabet{\mathcal}{usualmathcal}

\fancypagestyle{SPstyle}{
\fancyhf{}
\lhead{\colorbox{scipostblue}{\bf \color{white} ~SciPost Physics }}
\rhead{{\bf \color{scipostdeepblue} ~Submission }}

\fancyfoot[C]{\textbf{\thepage}}
}

\usepackage[utf8]{inputenc}
\usepackage{amsmath,amssymb,bbm}
\usepackage{graphicx}
\usepackage{xcolor}
\usepackage{mathtools}
\usepackage{float}
\usepackage{enumerate}

\newcommand{\ii}{\mathrm{i}}
\newcommand{\e}{\mathrm{e}}
\newcommand{\1}{\mathbbm{1}}
\newcommand{\ket}[1]{\left\lvert #1 \right\rangle}
\newcommand{\bra}[1]{\left\langle #1 \right\rvert}
\newcommand{\braket}[2]{\left\langle #1 \middle| #2 \right\rangle}
\newcommand{\mel}[3]{\left\langle #1 \middle| #2 \middle| #3 \right\rangle}
\newcommand{\ev}[1]{\left\langle #1 \right\rangle}
\newcommand{\abs}[1]{\left\lvert #1 \right\rvert}
\newcommand{\norm}[1]{\left\| #1 \right\|}
\newcommand{\dd}{\,\mathrm{d}}

\newcommand{\cmvM}{\mathcal M}
\newcommand{\cmvL}{\mathcal L}
\newcommand{\R}{\mathfrak{R}}
\renewcommand{\L}{\mathfrak{L}}
\newcommand{\F}{\mathfrak{F}}
\newcommand{\Fr}{R}
\newcommand{\Fl}{L}
\newcommand{\crc}{\mathbb{T}}
\newcommand{\T}{\mathrm{S}}
\newcommand{\M}{\mathrm{M}}
\newcommand{\ve}{\boldsymbol e}
\newcommand{\vep}{\boldsymbol e_+}
\newcommand{\vem}{\boldsymbol e_-}
\newcommand{\xopuc}{\eta}

\newcommand{\HH}{{\mathcal H}}
\newcommand{\LL}{{\mathcal L}}
\newcommand{\MM}{{\mathcal M}}
\def\S{{Szeg\H{o} }}
\def\nor#1{\|#1\|}

\newcommand{\diff}[1]{{#1}}

\begin{document}

\title{Solvable relaxation in discrete unitary systems: Ruelle-Pollicott resonances and CMV matrices}
\author{Urban Duh}
\author{Friedrich H\" ubner}
\author{Marko Žnidarič}
\date{\today}

\pagestyle{SPstyle}

\begin{center}{\Large \textbf{\color{scipostdeepblue}{
Solvable relaxation in discrete unitary systems: Ruelle-Pollicott resonances and CMV matrices\\
}}}\end{center}

\begin{center}\textbf{
Urban Duh\textsuperscript{1$\star$},
Friedrich H\" ubner\textsuperscript{2} and
Marko Žnidarič\textsuperscript{1}
}\end{center}

\begin{center}
{\bf 1} Physics Department, Faculty of Mathematics and Physics, University of Ljubljana, 1000 Ljubljana, Slovenia
\\
{\bf 2} Laboratoire de Physique de l’École Normale Superieure, CNRS, ENS \& Université PSL,
Sorbonne Université, Université Paris Cité, 75005 Paris, France
\\[\baselineskip]
$\star$ \href{mailto:urban.duh@fmf.uni-lj.si}{urban.duh@fmf.uni-lj.si}
\end{center}

\section*{\color{scipostdeepblue}{Abstract}}
\textbf{\boldmath{%
Leading eigenvalues of the truncated propagator, known as Ruelle-Pollicott (RP) resonances, are an elegant way of addressing the dynamics of unitary many-body systems. We study unitary propagators in their canonical form, known in the mathematical literature as the CMV matrices, and obtain a number of exact results for RP resonances and the associated norm-diverging eigenvectors. For the simplest CMV class describing a unilateral shift with an impurity, motivated by operator dynamics in dual-unitary circuits, we obtain closed-form results and in particular show that the three independent ways of obtaining RP resonances -- the truncated propagator, analytic continuation of the resolvent, and the rigged Hilbert space approach -- all give the same results. In more realistic CMV matrices, in which shift-like operator dynamics characteristic of chaotic systems is only asymptotic, we rely on the rich theory of orthogonal polynomials on the unit circle and identify two phases. In the first phase, relaxation occurs due to local operators effectively evolving into increasingly nonlocal ones with negligible backflow. Especially interesting is the second phase, which, surprisingly, exhibits faster relaxation because of contributions from the backflow of large operators. Additionally, in the second phase, RP resonances are not equal to the eigenvalues of the truncated propagator, instead, they are ``hidden'' within a ring of ill-conditioned eigenvalues.
}}

\vspace{\baselineskip}

%%%%%%%%%% BLOCK: Copyright information
% \noindent\textcolor{white!90!black}{%
% \fbox{\parbox{0.975\linewidth}{%
% \textcolor{white!40!black}{\begin{tabular}{lr}%
%   \begin{minipage}{0.6\textwidth}%
%     {\small Copyright attribution to authors. \newline
%     This work is a submission to SciPost Physics. \newline
%     License information to appear upon publication. \newline
%     Publication information to appear upon publication.}
%   \end{minipage} & \begin{minipage}{0.4\textwidth}
%     {\small Received Date \newline Accepted Date \newline Published Date}%
%   \end{minipage}
% \end{tabular}}
% }}
% }
%%%%%%%%%% BLOCK: Copyright information

% \linenumbers

% \vspace{10pt}
% \noindent\rule{\textwidth}{1pt}}
\pagebreak
\tableofcontents
% \noindent\rule{\textwidth}{1pt}
% \vspace{10pt}
\pagebreak

\section{Introduction}

The goal of the paper is to understand how relaxation arises within purely
unitary dynamics, that is, relaxation due to complicated, e.g., chaotic,
internal dynamics without any external dissipation or coupling. By relaxation, we
mean the decay of (auto)correlation functions $C(t):=\langle A(0)A(t) \rangle$,
where $A(t)$ is the time evolved observable $A$, and the
expectation value is with respect to some invariant state\footnote{A common
choice is the infinite temperature state $\rho \sim \1$, which is the one
we have in mind in this paper, although all approaches presented
work also for other invariant states.}.
Denoting by $U$ a one time-step unitary evolution propagator on the Hilbert
space $\HH$ of observables (sometimes called a superpropagator or a Heisenberg propagator), the time
evolved operator is simply $\ket{A(t)}=U^t \ket{A}$, and consequently 
\begin{equation}
  C(t)=\mel{A}{U^t}{A}.
  \label{eq:Ct}
\end{equation}
Provided $U$ is of finite dimension $N$, one can use its spectral decomposition,
\begin{equation}
  U=\sum_j \e^{\ii \omega_j} \ket{b_j}\bra{b_j},
  \label{eq:U}
\end{equation}
in terms of eigenvalues $\e^{\ii \omega_j}$ and eigenstates $\ket{b_j}$,
expressing the correlation function as a sum of oscillating terms,
\begin{equation}
  C(t)=\sum_j |c_j|^2 \e^{\ii \omega_j t},\qquad c_j=\braket{b_j}{A}.
  \label{eq:Cs}
\end{equation}
At first sight, it might seem that unitarity,
i.e., all eigenvalues of $U$ lying on the unit circle, is incompatible with
exponential relaxation (an eigenvalue inside the unit circle).

There is, however, no contradiction, one just has to carefully take the infinite
size limit $N \to \infty$ before sending $t \to \infty$. Indeed, there is no true relaxation
for finite $N$; one will eventually have non-physical recurrences
at times larger than the Heisenberg time, as signaled by unital eigenvalues of
$U$. However, when sending $N \to \infty$ first, the spectrum of $U$ can become
continuous and those finite-size effects are ``pushed to infinity''.
This shows that in order
to understand relaxation in the thermodynamic limit, relying on properties of
eigenvalues and eigenvectors may not be the best approach. While one might be
tempted to simply replace the sum in Eq.~\eqref{eq:U} by an integral, keeping
the ket and bra notation for right and left eigenvectors, functional analysis
tells us that for a $\e^{\ii \omega}$ from the continuous part of the spectrum
$\sigma_{\rm c}$, $\e^{\ii \omega} \in \sigma_{\rm c}$, there does not exist a
corresponding Hilbert space element $\ket{b} \in \HH$ such that $U
\ket{\omega}=\e^{\ii \omega}\ket{b}$. Behavior of infinite-dimensional
operators can be qualitatively richer than that of finite-dimensional ones,
and, for a good reason, functional analysis is required to deal with the former
while linear algebra suffices for the latter.

Let us briefly remind ourselves that the spectrum $\sigma$ of an operator $U$
can be split into three distinct subsets, $\sigma = \sigma_{\rm p} \cup \sigma_{\rm c}
\cup \sigma_{\rm r}$. Vaguely speaking, the point spectrum $\sigma_{\rm p}$ is
analogous to eigenvalues for finite $N$; for $z \in \sigma_{\rm p}$ the
resolvent $R(z)$ (Green's function)
\begin{equation}
  R(z) := (z-U)^{-1},
  \label{eq:R}
\end{equation}
does not exist because there exists an eigenvector $\ket{z} \in \HH$ satisfying
$U\ket{z}=z\ket{z}$  (causing $z-U$ to map different elements to zero). For the
continuous spectrum, $z \in \sigma_{\rm c}$, the resolvent $R(z)$ exists on a
dense set in $\HH$, however, it is unbounded and as a consequence one can only
find a series of approximate eigenvectors whose norm diverges and, therefore,
there is no true eigenvector in $\HH$. Thus, for a continuous spectrum, and our $U$ in
the thermodynamic limit will have a continuous spectrum on the unit circle,
strictly speaking, the ket and bra notation like in Eq.~\eqref{eq:U} does not
make sense (one can, though, define a projector to the eigensubspace). The
residual spectrum, $z \in \sigma_{\rm r}$, is a remaining part of the spectrum
for which the resolvent exists but is not defined on a dense set\footnote{For $z
\in {\mathbbm C}$ to be in the complement of the spectrum (the resolvent set)
three conditions are required: (i) $z-U$ is injective, (ii) $R(z)$ is bounded,
and (iii) $R(z)$ is defined on a dense set. If (i) does not hold $z \in
\sigma_{\rm p}$, if (ii) does not hold then $z \in \sigma_{\rm c}$, and if (iii)
does not hold then $z \in \sigma_{\rm r}$, see e.g. Ref.~\cite{Kreyszig} for
details.} (for normal operators $\sigma_{\rm r}$ is an empty set).

The question is, how to most easily identify and study the decay of correlation
functions without having to dwell too deeply into formal mathematics. An idea
that aims at that is known under the name of Ruelle-Pollicott
resonances~\cite{Ruelle,Pollicott,Gaspard}. If successful, it allows writing the
correlation function as a sum of only few terms (at long times even a single
one), like in Eq.~\eqref{eq:Cs}, though not in terms of the spectrum of $U$ but
rather in terms of the so-called Ruelle-Pollicott (RP) resonances. There are
different ways to get those and one of the aims of the present paper is to
explicitly demonstrate on solvable models that all these different approaches
give the same result while also discovering interesting physics and mathematics
along the way. To get a handle on relaxation, i.e., on RP resonances, there are
broadly speaking two approaches; one is motivated by physics, the other by
mathematics.

For the physics motivation, let us reflect back on the correlation function in
Eq.~\eqref{eq:Cs}: it is clear that even in an infinite-dimensional $\HH$ of a
chaotic system, not all observables $A$ will relax exponentially. In fact,
for locally-interacting many-body systems, e.g., a quantum circuit or a spin
chain, one in general expects decay only of sufficiently nice observables, for
instance, of local ones. For many, an extreme case would be an eigenspace
projector, there certainly will not be any relaxation. Therefore, from the point
of view of relaxation, the full Hilbert space $\HH$ is ``too big'' -- it contains
some observables that do relax, but many that do not. To bring out the
relaxation, one has to focus on the physically (experimentally) relevant
observables, for instance the local ones, and then one can expect to explicitly
see relaxation. This specifically suggests projecting an infinite-dimensional
$U$ to a subspace of observables with local support, obtaining the so-called
truncated propagator~\cite{prosenRuelleResonancesQuantum2002}. Due to the projection,
its spectrum is contained inside the unit circle and, provided the largest
eigenvalue $\lambda_1$ (largest in modulus $|\lambda_j|$) is isolated in an
appropriate limit, the decay of $C(t)$ at sufficiently large times should be
\begin{equation}
  C(t) \asymp |c_1|^2 \lambda_1^t,
\end{equation}
that is exponential decay if $|\lambda_1|<1$. Recently, the truncated propagator
method has been brought into focus again and used in different many-body
contexts~\cite{pre,xxz,xxz2,CA,prx,3site,duhCascade,arijeet,yamashika2026quantummanybodympembaeffect} (along with
the alternative RP approach using weak Lindbladian dephasing~\cite{moriLiouvilliangapAnalysisOpen2024, jacoby2025spectralGaps, zhang2026thermalizaionRates, yoshimuraTheoryIrreversibilityQuantum2025, duarte2026RP, mcculloch2026longlivedlocalquantumcoherences}), ranging from
quantum chaotic to integrable Floquet systems and classical cellular automata.

We have seen above that the spectrum of an operator can depend on the domain on
which it is defined, e.g., on the space $\HH$ of square summable functions
$\ell_2$, the spectrum is on the unit circle, while it is inside of it if defined
on the subspace of local observables. 
While this might seem surprising at first, there are many analogous more
familiar physical situations. For instance, solving the wave equation to find
eigenfrequencies of a metal plate, it is clear that those will depend on boundary
conditions. Boundary conditions select an appropriate set of functions -- the
operator domain -- and changing the domain can completely change the properties
of an (infinite-dimensional) operator\footnote{A simple textbook example is the
quantum mechanical momentum operator $p=-\ii \partial_x$ on an interval $x \in
[0,1]$. If one is in an infinite potential well, where fixed boundary conditions
$\psi(0)=\psi(1)=0$ are enforced, $p$ is not self-adjoint and does not have a
real spectrum; if one, on the other hand, has periodic boundary conditions
$\psi(x+1)=\psi(x)$ then $p$ is self-adjoint with a real
spectrum~\cite{surprises}.}.

A formalism that explicitly takes into account a subspace that is not
necessarily equal to $\HH$, is the so-called rigged (or equipped) Hilbert space
approach, where one needs a Gelfand triplet: in addition to the Hilbert space
$\HH$, one defines a subset of ``test functions'' $\Phi$ and a dual set of
linear functionals $\Phi^\times$. In our language, $\Phi$ is the
set of relevant observables $A$, for instance, a set of operators with strictly
locally supported densities, while $\Phi^\times$ is the set of bras and kets used in
calculating expectation values of $A$. Because $\Phi$ is a subset of $\HH$
containing more ``tame'' observables than the full $\HH$, the set of $\Phi^\times$ is
larger than $\HH$ and contains also more ``wild'' objects than $\HH$ (the so-called distributions),
i.e., $\Phi \subset \HH \subset \Phi^\times$. The rigged Hilbert space formalism
has been introduced in the '60~\cite{gelfandGeneralizedFunctionsApplications1964},
among other to formalize Dirac's bra and ket notation to operators with a
continuous spectrum, for instance, the position operator in 1D whose generalized
eigenvectors are Dirac delta functions and are not from $L^2(\mathbbm{R})$ but
rather from $\Phi^\times$~\cite{delamadridRoleRiggedHilbert2005} (remember that
for $z \in \sigma_{\rm c}$ there are no eigenvectors in $\HH$). Within the
rigged Hilbert space, one can prove a generalized spectral theorem so that a
decomposition like in Eq.~\eqref{eq:U} holds in $\Phi^\times$ also for operators
with a continuous spectrum\footnote{One can remark that while the standard
functional analysis (without rigged Hilbert spaces) suffices, Dirac's notation
is rather convenient, or, as mentioned on p.~244 of Ref.~\cite{reedMathPhysI}, we
physicists are emotionally attached to it.}. While few introductory quantum
mechanics textbooks mention it~\cite{Ballentine} because physicists have learned
to live without it, it does help to understand, e.g., why some operations are not
``permitted'' on Dirac delta functions~\cite{surprises}. The formalism has been
mostly used in
studies of decay~\cite{darius}, scattering~\cite{deMadrid2002continuous, deMadrid2004rectangular},
irreversibility~\cite{prigogine}, and in the '90 also in the context of RP
resonances of classical chaotic
maps~\cite{hasegawaUnitarityIrreversibilityChaotic1992,Antoniou}. 
% We shall show
% that in our exactly solvable cases, the rigged Hilbert space approach gives
% exactly the same RP resonances as the truncated propagator method.

The last approach that we are going to use to extract the RP resonances is to
analytically continue the resolvent $R(z)$ for the full unitary $U$ from the
outside of the unit circle, where it is analytic, past the cut at the
unit circle to the inside of it, finding isolated poles (resonance) at $|z|<1$.
This was, in fact, the approach already discussed in the earliest works like
Ref.~\cite{hasegawaUnitarityIrreversibilityChaotic1992}. One of the merits of
our paper is that, in the simple solvable models that we study, one can
explicitly show that the resolvent method as well as the rigged Hilbert space
approach -- two methods that usually cannot be executed in more complicated
models -- give the same result as the truncated propagator method, offering a
conceptual and mathematical foundation for a typically numerical approach. The
truncated propagator method can, therefore, be though of as a
tool by which one can address many questions about dynamics of many-body systems
without having to worry about subtle details of functional analysis or rigged
Hilbert spaces connected to infinities.

Our work aims to contribute to a scarce collection of models where RP
resonances can be computed exactly. Those consist mostly of older works on classical models~\cite{hasegawaUnitarityIrreversibilityChaotic1992,
Antoniou, prigogine, Gaspard, chaosBook, antoniou1997generalizedSpectral,
gaspard1995spectralSignature}, and few more recent quantum many-body models with disorder~\cite{yoshimuraTheoryIrreversibilityQuantum2025,jacoby2025spectralGaps} that can be solved also by other means. 
% Here we focus on models solvable in the truncated propagator approach.
What are the solvable models for which we will be able to execute all three
methods? Little known in physics, \diff{see though recent Refs.~\cite{yehPRB25,trunin25,yehPRB26}}, but much more studied in mathematics, are
the so-called CMV matrices~\cite{simonOrthogonalPolynomialsUnit2005,golinskii}
which, it turns out, are generic forms all unitary operators have in an
appropriate basis (with some technical conditions to be defined). A lot of
beautiful mathematics and connections across several fields have been established
in the last 20 years using CMV matrices, and one of our aims is also to bring some of
those results to the attention of physicists. 
Let us have a brief look at what CMV matrices are and why they are important.

% In addition, we also reveal an
% interesting phase diagram that a class of such CMV matrices exhibits, including
% a phase in which a ring of spurious badly conditioned eigenvalues shields a
% hidden leading RP resonance inside the ring. In such a case, a naive truncation
% will not correctly identify the leading RP resonance, however, we provide a
% remedy. In addition, we also identify a case of non-exponential Gaussian decay
% of correlation functions. 

\subsection{CMV matrices}
\label{sec:cmv_intro}

CMV matrices\footnote{Named after Cantero, Moral and Vel\' azques who in their
paper~\cite{cmv} made an important contribution, see preface in
Ref.~\cite{simonOrthogonalPolynomialsUnit2005} for the historical background.} are
unitary matrices with a simple pentadiagonal structure. Any unitary
operator\footnote{\label{ft:cyclicity}Strictly speaking, any $U$ that has a bicyclic vector $x_0$ can
be brought to the CMV form. A bicyclic vector is a vector such that $U^t x_0,
t=\dots, -2, -1, 0,1,2,\ldots$ forms a basis, i.e., linear span is dense (for a cyclic vector, one would take only positive powers). Operators without a
bicyclic vector can be written as a direct sum of ones that do have a bicyclic
vector. For finite $N$, a bicyclic property is the same as a non-degeneracy (simple
spectrum).
% \M{smels like absence of symmetries?} \U{It does. Would be interesting if we 
% could cook up a CMV for $U(1)$ circuits and see transport.}
} can be brought to such a
form in an appropriate basis. A CMV matrix $U$ can be parametrized by complex
coefficients $\alpha_k,\, k=0,1,2,\ldots$, called Verblunsky coefficients, and
can be written as a product of two block-diagonal matrices, 
\begin{align}
  \label{eq:cmv}
    U &:= \LL \MM, \\
    \LL &:= \begin{pmatrix}
    \Theta_0 & \\
             & \!\! \Theta_2 \\
    && \!\!\ddots
    \end{pmatrix}\!=\Theta_0 \oplus \Theta_2 \oplus \cdots , &\MM &:= \begin{pmatrix}
        \1_{1 \times 1} & \\
        & \!\!\!\Theta_1 \\
        && \!\!\ddots
    \end{pmatrix}\!=\mathbbm{1}_{1\times 1} \oplus \Theta_1 \oplus \cdots,\\ 
    \Theta_k &:= \begin{pmatrix}
        \alpha_k^* & \rho_k \\
        \rho_k & - \alpha_k
    \end{pmatrix}, \qquad &\rho_k &:= \sqrt{1 - \abs{\alpha_k}^2},
\end{align}
where $\mathbbm{1}_{1 \times 1}$ is a $1\times 1$ identity matrix, i.e., a
single $1$, while all $\Theta_k$ are $2 \times 2$ matrices. Written explicitly,
\begin{equation}
    U = \begin{pmatrix}
        \alpha_0^* & \alpha_1^* \rho_0 & \rho_1 \rho_0 & 0 & 0 & \dots \\
        \rho_0 & - \alpha_1^* \alpha_0 & -\rho_1 \alpha_0 & 0 & 0 & \dots \\
        0 & \alpha_2^* \rho_1 & - \alpha_2^* \alpha_1 & \alpha_3^* \rho_2 & \rho_3 \rho_2 & \dots \\
        0 & \rho_2 \rho_1 & -\rho_2 \alpha_1 & -\alpha_3^* \alpha_2 & - \rho_3 \alpha_2 & \dots \\
        0 & 0 & 0 & \alpha_4^*\rho_3 & - \alpha_4^* \alpha_3 & \dots \\
        \vdots & \vdots & \vdots & \vdots & \vdots & \ddots
    \end{pmatrix}.
\end{equation}
Such $U$ is a
unitary operator on $\HH=\ell_2:= \left\{\sum_{n = 0}^\infty c_n \ket{n}
; \ \sum_{n = 0}^\infty \abs{c_n}^2 < \infty \right\}$. Note that the
matrix element $\mathbbm{1}_{1\times 1}$ in ${\cal M}$ is what breaks a $2\times
2$ block diagonal structure and makes $U$ interesting, also causing, as we will
see, strong even-odd effects in eigenvectors.

To appreciate the role that CMV matrices have among unitary operators, let us
start with perhaps more familiar self-adjoint operators. Taking a self-adjoint
$A$ and a cyclic vector $x_0$, we can form a basis from $x_0,Ax_0,A^2x_0,\ldots$.
Applying the Gram-Schmidt orthogonalization, one gets an orthonormal basis
$x_j=p_j(A)x_0$, where $p_j(A)$ is a polynomial of degree $j$. In this basis, $A$
is a tridiagonal matrix $J$, $\mel{x_j}{A}{x_k}=0$ for $|j-k|>2$, called a
Jacobi matrix parametrized by the diagonal elements and the above-diagonal ones.
In the context of Heisenberg evolution of Hamiltonian systems, it is often called a Krylov or a Lanczos
basis, being rather popular lately~\cite{Dimarsky}. Polynomials $p_j$ satisfy a
three-term recursion relation and are orthogonal under an appropriate
measure\footnote{Two of the simplest and most famous examples are Hermite
polynomials for a Gaussian measure on $\mathbbm{R}$, and Legendre polynomials
for a uniform measure on the interval $[-1,1]$.} $\dd\mu$ on the real line
$\mathbbm{R}$. Moments of the measure are equal to the powers of $A$, $\int\!
x^t \dd\mu(x)=\mel{x_0}{A^t}{x_0}$. There is a one-to-one correspondence between a
self-adjoint $A$, its tridiagonal Jacobi form $J$, the measure $\dd\mu$, and
orthogonal polynomials $p_j(x)$ on the real line. Understanding self-adjoint
operators can be done in any of those equivalent settings, whichever one is the
simplest for the problem in question. Needless to say, orthogonal polynomials on
the real line connect many different areas of mathematics~\cite{orto}.

One can wonder if there also exists a similar simple form that unitary operators $U$
take in an appropriate basis. If one tries to form a basis by orthonormalizing
% $\{1,z,z^2,\ldots\}$,
$\{x_0, U x_0, U^2x_0, \dots\}$
in analogy with Jacobi matrices, it does not work -- one
does not end up with a banded representation of $U$. The insight of
CMV~\cite{cmv} was to orthonormalize
% $\{1,z,z^{-1},z^2,z^{-2},\ldots \}$,
$\{x_0,Ux_0,U^{-1}x_0,U^2x_0,$ $U^{-2}x_0,\ldots \}$,
in which case one does obtain a 5-diagonal\footnote{Which is optimal.} form of $U$
written in Eq.~\eqref{eq:cmv}. Verblunsky coefficients determine a 3-term
recurrence relation for polynomials $\Phi_j$ that are orthogonal on the unit
circle $\crc :=\{z;\, |z|=1\}$ under a measure $\dd\mu(\crc)$.
Again, there is a one-to-one correspondence between $U$, a 5-diagonal CMV matrix
with its Verblunsky coefficients $\alpha_k$, orthogonal polynomials on the unit
circle (OPUC), and the measure $\dd\mu$. While the beginnings of OPUC go back a
century to Szeg\H{o}'s pioneering work, the field received fresh impetus with the CMV
paper~\cite{cmv}. As we will see, OPUC are connected with spectral theory,
complex analysis and more. Standard references on the subject are
books~\cite{simonOrthogonalPolynomialsUnit2005,simonOrthogonalPolynomialsUnit2005a},
with shorter overviews also in Refs.~\cite{simon_cmv5,golinskii,golinskiitotik}.
We also mention that CMV matrices can be connected to integrable systems,
specifically to the Ablowitz-Ladik model (discrete nonlinear Schr\" odinger
equation)~\cite{nenciu}. More details will be given in Section~\ref{sec:opuc},
here let us just highlight a few results most relevant for relaxation.  

In our context of correlation functions decay, the space on which CMV
matrices act is the space of operators, and the correlation function in
Eq.~\eqref{eq:Ct} is equal to the $t$-th moment of the measure,
\begin{equation}
  C(t)=\mel{x_0}{U^t}{x_0}=\ev{z^t, 1}_\mu=\int_\crc {z^{-t} \dd\mu(z)}=\int_{0}^{2\pi}{\!\e^{-\ii t \theta}w(\theta)\dd\theta},
  \label{eq:Cmu}
\end{equation}
where, in the last equality, we wrote the measure $\dd\mu$ explicitly in terms of
its density\footnote{The measure can be in general complicated, however, in all
the CMV cases we study it is a nonzero continuous function on $\mathbbm{T}$ and
can therefore be expressed in terms of a simple density $w$.} $w(\theta)$
parametrizing the unit circle by $\theta$. Therefore, one can approach the question of
the asymptotic decay of $C(t)$ via the so-called moments problem -- a field of
mathematics studying the properties of measures. Looking at the last equality in
Eq.~\eqref{eq:Cmu}, we also recognize that one deals with a Fourier transformation
of $w$, which is exactly how Toeplitz operators (matrices) are defined. Namely,
a compact way of defining an infinite Toeplitz matrix $T$ in terms of its
symbol\footnote{A physicist might call $f(\e^{\ii k})=H(k)$ a Bloch
Hamiltonian.} $f(z)$ is $T_{j,k}=\int f(\e^{\ii \theta})\e^{-\ii
(k-j)\theta}\dd\theta$. The behavior of $C(t)$ is therefore connected to certain
Toeplitz operators with a non-negative symbol and the asymptotic decay of $C(t)$ to the
asymptotics of Toeplitz matrices, which is a central object in a number of fine
results in theoretical physics, e.g., the exact solution of the 2D Ising
model~\cite{toeplitz}. In the truncated propagator approach to RP resonances, one
can truncate a CMV matrix by simply taking the upper-left $N\times N$ block of
an infinite $U$, obtaining a finite truncation $U_N$. Eigenvalues of such
sub-unitary $U_N$ should then give us the RP resonances.

A nice result about CMVs is~\cite{simon_cmv5} that the characteristic polynomial
of such $U_N$ is nothing but the OPUC $\Phi_N$,
\begin{equation}
  \det{(z-U_N)}=\Phi_N(z).
\end{equation}
The leading RP resonance $\lambda_1$ should therefore be equal to the largest
zero $z_j$ of $\Phi_N$. While true in a large class of models, we also find an interesting
example where this is not the case and one has to use another analytical
trick to get the resonance.

We think that CMV matrices could be of wider use in physics, not just in the
context of RP resonances. Few previous cases where matrices of
such form have appeared are in the Krylov context in \diff{Refs.~\cite{aditi, suchsland2025krylov,yehPRB24,yehPRB25,yehPRB26,yehAnderson,trunin25}.} OPUC have
been used for mutually unbiased bases useful in quantum
information~\cite{simanek}, and CMV matrices are also mentioned in
Ref.~\cite{toeplitz}. \diff{One can also show that the operator (Heisenberg) propagator of a semi-infinite kicked transverse field Ising model (or, equivalently, the kicked XY model) has an exact CMV form in the majorana basis~\cite{yehPRB24,yehPRB25,yehPRB26,trunin25}. This means that there is a mapping between Verblunsky coefficients and a particular noninteracting model with inhomogeneous parameters. Note that our motivation for using CMV matrices is different -- we are using them as an effective description of operator dynamics (coarse-grained according to their support) in homogeneous infinite systems.}

\subsection{Summary of results}

The simplest nontrivial case\footnote{The simplest CMV matrix is the one with
$\alpha_k=0$, the so-called free case, for which the correlations decay
immediately, $C(t)=\delta_{t,0}$.} we study is that of CMV matrices with a single
nonzero $\alpha_0$ (Sec.~\ref{sec:shift}). The dynamics they describe is similar
to the operator dynamics found in dual-unitary circuits, if we
interpret the CMV state $\ket{n}$ as a coarse-graining of operators of
support proportional to $n$. We determine the
location of the single RP resonance in three ways: (i) by studying the
poles of the resolvent on the second Riemann sheet (i.e., of its analytic
continuation from outside the unit circle to the inside) in
Sec.~\ref{sec:shift_res}, (ii) through the analytic properties of generalized
eigenvectors and the generalized spectral measure provided by an analog of the
spectral theorem in the rigged Hilbert space in Sec.~\ref{sec:shift_rigged} and
(iii) from the leading eigenvalue of the truncated propagator in
Sec.~\ref{sec:shift_tp}. All three approaches lead to the same result
$\lambda=\alpha_0$ and also provide us with RP eigenvectors $\ket{\R}, \bra{\L}$
(called Gamow vectors in the rigged Hilbert space formalism). The components of
Gamow vectors diverge as $\vline\braket{n}{\R}\vline^2 \sim \lambda^{-n}$ with
the ``operator support'' $n$ and are formally distributions from the dual space
$\Phi^\times$ constructed over the space of local observables $\Phi$. This means that
we obtain an effective non-unitary description for local observables at long
times, $\mel{\psi}{U^t}{\phi} \asymp \lambda^t \braket{\psi}{\R}
\braket{\L}{\phi}$ for any $\psi, \phi \in \Phi$.

\begin{figure}[ht]
    \centering
    \includegraphics[width=0.45\linewidth]{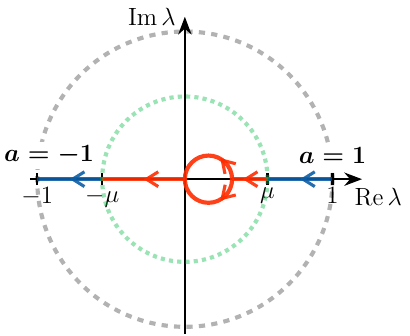}
    \caption{Location of the RP resonance $\lambda$ in the complex plane (full
    curves) as $a$ changes from $1$ (where $\lambda=1$) to $-1$ ($\lambda=-1$) for a fixed $\mu$ with
    arrows denoting the direction of decreasing $a$. The phase transition
    between phase I (blue) and phase II (red) occurs at $a = 0$ and at a negative value of 
    $a$ nontrivially depending on $\mu$, see also Sec.~\ref{sec:phases} and
    Figs.~\ref{fig:presek}, \ref{fig:fazni}.}
    \label{fig:rpr_diagram}
\end{figure}

In Sec.~\ref{sec:opuc} we then review the standard OPUC
results~\cite{simonOrthogonalPolynomialsUnit2005,simonOrthogonalPolynomialsUnit2005a}
that we use in Sec.~\ref{sec:cmvexp} to study CMV matrices with exponentially
decaying Verblunsky coefficients $\alpha_k=a \mu^{k}$. Such a choice implies
that the coupling between the forward and the backward flow of operators becomes
exponentially small, suggesting asymptotic shift-like dynamics. It is
motivated by Ref.~\cite{duhCascade}, where we argued that generic chaotic
operator dynamics tends towards shift-like dynamics with increasing operator
support. We find rich behavior that should be of interest to physicists and
mathematicians alike. Namely, fixing $\mu$ and varying $a$, the leading RP
resonance $\lambda$, i.e., the decay rate of correlations, continuously changes
from no decay with $\lambda= 1$ to a Gaussian decay at $\lambda=0$ and back to
no decay at $\lambda = -1$ (Fig.~\ref{fig:rpr_diagram}), however, other
properties exhibit a series of discontinuous changes. We identify two main
phases. In phase I, the leading eigenvalue of the truncated propagator $U_N$
indeed converges to the leading RP resonance as $N \to \infty$, while all the
other eigenvalues converge to a smaller ring with radius $\abs{z} = \mu <
|\lambda|$. The transition to phase II occurs when the leading eigenvalue
merges into the ring $\abs{z} = \mu$, seemingly disappearing, and correlations
decay faster than $\sim \mu^t$, i.e., $|\lambda|<\mu$. The leading eigenvalue of
$U_N$, being of modulus $\mu$, is therefore not equal to $\lambda$. To
nevertheless obtain $\lambda$, we present two methods. The first approach
(Sec.~\ref{sec:phases}) relies on the theorems from the OPUC literature, which
relate the leading RP resonance to the convergence radii of \S and Carath\'
eodory functions, both expressible as a limit of OPUC of increasing order.
Accounting for a subtle behavior of the limit, one gets the correct
$|\lambda| < \mu$. In the second approach (Sec.~\ref{sec:transfer}) we use the
transfer matrix method to derive (i) eigenvectors of the truncated propagator
and express them via OPUC and (ii) an exact expression for the resolvent. To obtain the \diff{leading} RP resonance \diff{in phase II} we then use numerical analytic continuation of the resolvent to the second Riemann sheet. \diff{Moreover, such analytical continuation also reveals subleading resonances in both phases I and II that are ``hidden'' within the ring, i.e., do not appear among the eigenvalues of the truncated $U_N$. We also obtain the corresponding Gamow vectors, showing that their components} asymptotically diverge as $\vline\braket{n}{\R_i}\vline^2
\sim\abs{\lambda_i}^{-n}$ (as in the simpler case), same as their
condition numbers $\kappa_i \sim |\lambda_i|^{-n}$, satisfying the conjectured
asymptotic equality from Ref.~\cite{duhCascade}. Furthermore, partial
binorms, which are important for the convergence of RP resonances in the truncated
propagator approach, are convergent in phase I but divergent in phase II.
Compatible with that, one also observes strong finite-size effects in phase II
(Sec.~\ref{sec:finite_size}).

\section{Shift with an impurity}
\label{sec:shift}

\begin{figure}[ht]
    \centering
    \includegraphics[width=0.9\linewidth]{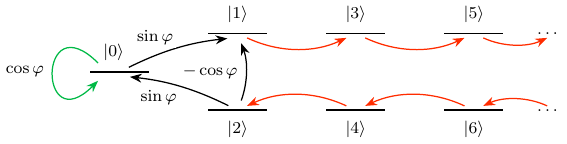}
    \caption{Diagram of the simplest nontrivial CMV model, $\alpha_0 = \cos
\varphi$ and $\alpha_{k  > 0} = 0$. Arrows denote nonzero matrix elements of the
propagator $U$.}
    \label{fig:shift_imp_diagram}
\end{figure}

We first focus on the simplest nontrivial example, where $\alpha_0 = \cos
\varphi, \varphi \in (0, \pi)$ and $\alpha_{k  > 0} = 0$. In this case,
the propagator can be written as
\begin{equation}
   U = \begin{pmatrix}
        \ket{0} & \ket{1}
    \end{pmatrix} \begin{pmatrix}
        \cos \varphi & \sin \varphi \\ \sin \varphi & -\cos \varphi
    \end{pmatrix} \begin{pmatrix}
        \bra{0} \\ \bra{2}
    \end{pmatrix} + \sum_{n = 1}^\infty \left(\ket{2n}\bra{2n + 2} + \ket{2n + 1} \bra{2n-1}\right) \label{eq:shift_imp}
\end{equation}
and can be interpreted as a scattering on a single impurity localized around
the $\ket{0}$ state analogous to Friedrich's model~\cite{gadellaFriedrichsModelIts2011}, as depicted in
Fig.~\ref{fig:shift_imp_diagram}. In this interpretation, even states $\ket{2n}$ correspond to ingoing particles and odd states $\ket{2n + 1}$ correspond to
outgoing particles. We instead adopt a many-body interpretation, where a pair of
even-odd states $\ket{2n}, \ket{2n + 1}$ is understood as a coarse-graining of
operators with support $n$. The considered model, therefore, has non-trivial
dynamics at support $1$ and trivial shift-like dynamics at all higher supports.
As shown in Ref.~\cite{duhCascade}, this corresponds to a particularly simple
example of dual-unitary circuits~\cite{bertini2019exactCorrelation,
bertini2026exactlySolvable}, which exhibit exact shift-like dynamics at high
supports, \diff{see also comments in Ref.~\cite{suchsland2025krylov}.} 

We are interested in behavior of correlation functions $\mel{n}{U^t}{m}$ for
large times $t$. Due to the simplicity of the considered model, one can
analytically check (see Appendix~\ref{app:shift_imp_c}) that they are
asymptotically either $0$ or decay as $(\cos \varphi)^t$ (\diff{the simplest one $n = m = 0$ has been briefly discussed in Ref.~\cite{yehPRB25}}). In the following, we
shall derive this via the approach of RP resonances in two different ways:
through the analytic continuation of the resolvent in Sec.~\ref{sec:shift_res}
and by analytic continuation of the generalized eigenvectors defined in the
context of rigged Hilbert spaces in Sec.~\ref{sec:shift_rigged}, and finally
show that the same result is obtained via the truncated propagator in
Sec.~\ref{sec:shift_tp}.

\subsection{Resolvent formalism}
\label{sec:shift_res}

We first extract RP resonances through analytic properties of the resolvent
closely following Ref.~\cite{hasegawaUnitarityIrreversibilityChaotic1992}.
The central result we use is the so-called resolvent identity
\begin{equation}
    U^t = \frac{1}{2\pi \ii} \oint_{\mathcal C} z^t R(z) \dd{z}, \label{eq:res_id}
\end{equation}
where $R(z) := (z - U)^{-1}$ is the resolvent, $\mathcal C$ is a positively
oriented curve just outside the unit circle (or, more generally, a curve encircling
the entire spectrum of $U$ once) and $t$ is a positive integer.

\begin{figure}[ht]
    \centering
    \includegraphics[width=0.45\linewidth]{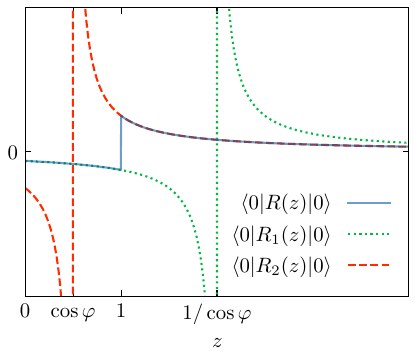}
    \includegraphics[width=0.54\linewidth]{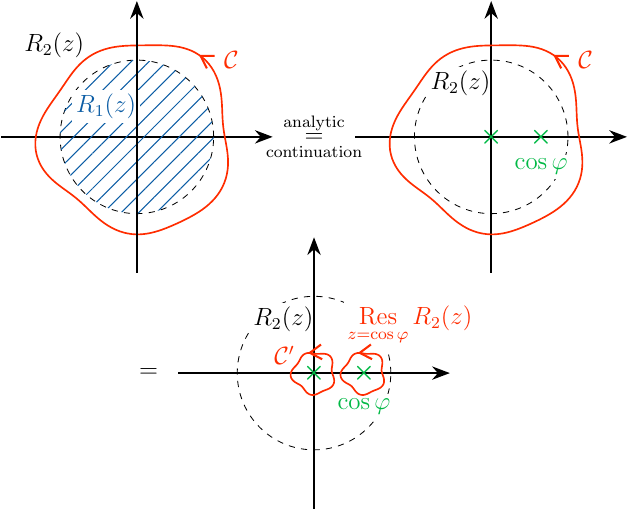}
    \caption{(Left) The resolvent $\mel{0}{R(z)}{0}$ in the simplest nontrivial
    CMV model, $\alpha_0 = \cos\varphi$ and $\alpha_{k  > 0} =
    0$, and its analytic continuations to the first ($R_1$, green) and second ($R_2$, red)
    Riemann sheet. (Right) Extraction of RP resonances within the resolvent
    formalism, diagram of the top line of Eq.~\eqref{eq:shift_rp_exp}. Red curves
    denote the integration contour in the complex plane, green crosses correspond to
    the poles of the integrand and dashed black curve is the unit circle.}
    \label{fig:shift_imp_resolvent}
\end{figure}

In the considered model \eqref{eq:shift_imp}, the resolvent can be obtained
exactly, either via the solution of the inhomogeneous eigenequation (see
Sec.~\ref{sec:cmv_resolvent} for details about this approach for a generic CMV
matrix) or by expanding it into a Neumann series $R(z) = \frac{1}{z} (1 -
U/z)^{-1} =$ $\frac{1}{z} \sum_{t = 0}^\infty U^t/z^t$ and using the exact
results for correlation functions in Appendix~\ref{app:shift_imp_c}.
Without going into details, we state the result
\begin{align}
    R(z) &= \ket{\R(z)} f(z) \bra{\L(z)} + \Gamma(z), \label{eq:shift_res} \\
    f(z) &:= \begin{cases}
        \frac{z + \cos \varphi}{2z(z - \cos \varphi)}; & \abs{z} > 1 \\
        -\frac{1/z + \cos \varphi}{2z(1/z - \cos \varphi)}; & \abs{z} < 1
    \end{cases}, \\
    \ket{\R(z)} &:= \ket{0} + \frac{1}{\sin \varphi} \left[\sum_{n = 2, 4, \dots}^\infty z^{n/2 - 1} (z - \cos \varphi) \ket{n} + \sum_{n= 1, 3, \dots}^\infty \frac{1 - z \cos \varphi}{z^{(n + 1)/2}} \ket{n} \right],  \label{eq:shift_res_r}\\
    \bra{\L(z)} &:= \bra{0} + \frac{1}{\sin \varphi} \left[ \sum_{m = 2, 4, \dots}^\infty \frac{1 - z \cos \varphi}{z^{m/2}} \bra{m} + \sum_{m = 1, 3, \dots}^\infty z^{(m - 1)/2} (z - \cos \varphi) \bra{m} \right],
\end{align}
where $\Gamma(z)$ is a continuous matrix-valued function with no poles for $0 <
\abs{z} < \infty$ (see Eq.~\eqref{eq:cmv_resolvent} for the general expression).
The exact expression for the simplest matrix element is $\mel{0}{R(z)}{0} = \begin{cases}
        \frac{1}{z - \cos\varphi}; & \abs{z} > 1\\
        -\frac{\cos\varphi}{1 - z \cos\varphi}; & \abs{z} < 1
    \end{cases}$ and is shown in Fig.~\ref{fig:shift_imp_resolvent}. It
contains the qualitative features that are also present for all other matrix
elements: it is an analytic function without poles for $\abs{z} > 1$ and
for $\abs{z} < 1$, and it has a discontinuity at $\abs{z} = 1$, i.e., along
its spectrum.

We now turn to evaluating Eq.~\eqref{eq:res_id}. Since the contour $\mathcal C$
lies strictly outside the unit circle, this can be done by analytically
continuing the expression for the resolvent \eqref{eq:shift_res} from $\abs{z} >
1$ to $\abs{z} \leq 1$. To be more precise, the resolvent outside the unit
circle $R_2(z) := R(\abs{z} > 1)$ is an analytic function that can be
analytically continued to the entire complex plane, obtaining what is usually
referred to as the resolvent on the second Riemann
sheet~\cite{hasegawaUnitarityIrreversibilityChaotic1992}\footnote{One can obtain
the resolvent on the first Riemann sheet by analytically continuing $R_1(z) :=
R(\abs{z} < 1)$ to the entire complex plane}. In the present case, performing
the analytic continuation is simple, one can simply use the expression $f(z)
= \frac{z + \cos \varphi}{2z(z - \cos \varphi)}$ valid for $\abs{z} > 1$.
$\mel{0}{R_2(z)}{0}$ is shown in Fig.~\ref{fig:shift_imp_resolvent}

It is now also easy to see that $R_2(z)$ has two poles\footnote{More precisely,
our statements should be understood for each matrix element of the resolvent
separately. All of them can only have poles at the specified locations.} within
the contour of integration: at $z = \cos \varphi$ and at $z = 0$. Since the pole
at $z = \cos \varphi$ is simple, its contribution can be evaluated using the
residue theorem and the contour of integration shrunk to $\mathcal C'$, an
arbitrarily small positively oriented contour encircling $z = 0$. The entire
procedure is shown diagrammatically in
Fig.~\ref{fig:shift_imp_resolvent}, writing it in equation
\begin{align}
    U^t &= \frac{1}{2\pi \ii} \oint_{\mathcal C} z^t R(z) \dd{z} = \frac{1}{2\pi \ii} \oint_{\mathcal C} z^t R_2(z) \dd{z}
    = (\cos \varphi)^t \mathop{\text{Res}}\limits_{z = \cos\varphi} R_2(z) +  \frac{1}{2\pi \ii} \oint_{\mathcal C'} z^t R_2(z) \dd{z} \nonumber \\
    &= (\cos \varphi)^t  \ket{\R(\cos \varphi)} \bra{\L(\cos \varphi)} + \xi, \label{eq:shift_rp_exp}
\end{align}
where $\text{Res}_{z = x}f(z) := \lim_{z \to x} (z - x) f(z)$ for a simple pole
and $\xi$ denotes the (potential) contributions to the integral from the
pole at $z = 0$.

Interpreting Eq.~\eqref{eq:shift_rp_exp}, the considered system has an RP resonance
$\lambda = \cos \varphi$ and the corresponding right/left RP ``eigenvectors''
\begin{align}
    \ket{\R} &:= \ket{\R(\cos \varphi)} = \ket{0} + \sin \varphi \sum_{n = 1, 3, \dots}^\infty (\cos \varphi)^{-(n + 1)/2} \ket{n}, \label{eq:shift_rp_r} \\
    \bra{\L} &:= \bra{\L(\cos \varphi)} = \bra{0} + \sin \varphi \sum_{m = 2, 4, \dots}^\infty (\cos \varphi)^{-m/2} \bra{m} \label{eq:shift_rp_l}
\end{align}
respectively. Note that the ``eigenvectors'' $\ket{\R}$ and $\bra{\L}$ are not normalizable and furthermore $\bra{\L}\neq \ket{\R}^\dagger$. We will refer to $\ket{\R}, \bra{\L}$ as the right, left Gamow
vectors, analogous to the language used in scattering theory, where
they describe exponentially decaying states~\cite{bohmDiracKetsGamow1969}. In our case,
Gamow vectors give us an effective dissipative description of dynamics.
Evaluating Eq.~\eqref{eq:shift_rp_exp} at arbitrary basis
elements, we can show that the contributions from $(\cdots)$ are zero for large
enough $t$. We have thus shown that the correlation functions in our system
asymptotically indeed either decay as $\mel{n}{U^t}{m} \sim (\cos \varphi)^t$ or vanish identically after finite time\footnote{$\mel{n}{U^t}{m} = 0$ for large enough $t$ if either $\braket{n}{\R} = 0$
or $\braket{\L}{m} = 0$.} $t$.

The prefactor of the decay is encoded in Gamow vectors, which have a peculiar
structure. Namely, their components diverge exponentially as $\sim (\cos
\varphi)^{-n/2}$ and exhibit a strong odd-even staggering. This is compatible
with the observations for dual-unitary circuits in Ref.~\cite{duhCascade}, an
interpretation of $\ket{\R}$, $\bra{\L}$ being lattice analogs of unstable and
stable manifolds (iterations forward and backward in time that tends towards
either odd $\ket{2n + 1}$ or even $\ket{2n}$ states, respectively,
cf.~Fig.~\ref{fig:shift_imp_diagram}) of classical chaotic systems, and the
structure of CMV matrices that together book-keeps $z^j$ and $z^{-j}$. However,
it also implies that both $\ket{\R}$ and $\ket{\L}$ are not elements of $\ell_2$
and thus do not have a well-defined inner product with an arbitrary element of
$\ell_2$. In the following section, we shall clarify this in the
rigged Hilbert space formalism.

\subsection{Rigged Hilbert space formalism}
\label{sec:shift_rigged}

In this section, we shall introduce rigged Hilbert spaces, which will allow us
to work with vectors that are ``too wild'' to be elements of $\ell_2$, such as
the Gamov vectors $\ket{\R}$ and $\bra{\L}$ in Eq.~\eqref{eq:shift_rp_exp}. The intuitive idea of rigged Hilbert spaces is to
restrict ourselves to a subset $\Phi$ of the considered Hilbert space consisting
of ``tame enough'' objects which have a well-defined inner product with the
diverging Gamow vectors we wish to describe. The rigged Hilbert space
formulation will also allow us to recover a variant of the spectral
decomposition with (generalized) eigenvectors of our model~\eqref{eq:shift_imp},
which has a purely continuous spectrum and thus does not have any eigenvectors
in $\ell_2$. Analytic properties of generalized eigenvectors will then allow us
to derive the RP resonances description equivalent to the one in
Sec.~\ref{sec:shift_res}.

The notion of rigged Hilbert is not foreign in physics. One of the most widely
used examples of the rigged Hilbert space construction are generalized functions
(distributions)~\cite{reedMathPhysI}. They can be constructed by restricting the
Hilbert space of square-integrable functions $L^2(\mathbb{R})$ to the space of
rapidly decaying smooth functions (Schwartz functions). This then allows us to
work with functionals representing either locally singular objects (e.g., Dirac
deltas) or extended objects (e.g., plane waves used in Fourier transform).

\subsubsection{Constructing the rigged Hilbert space}

Let's now be more precise. Let $\mathcal H$ be a Hilbert space and $\Phi \subset
\mathcal H$ be its dense subset. The two spaces, together with the dual space
$\Phi^\times$ of linear functionals over $\Phi$, constitute a rigged Hilbert
space or a Gelfand triple
\begin{equation}
    \Phi \subset \mathcal H \subset \Phi^\times.
\end{equation}
For a more rigorous introduction to rigged Hilbert spaces see
Refs.~\cite{delamadridRoleRiggedHilbert2005, bohmDiracKetsGamow1969,
gelfandGeneralizedFunctionsApplications1964}.

In the model~\eqref{eq:shift_imp}, we have $\mathcal H = \ell_2$. There are
multiple possible choices of $\Phi$, the choice that will be appropriate in the
present case is
\begin{equation}
    \Phi = c_{00} := \left\{\sum_{n = 0}^\infty c_n \ket{n}; \ c_n \neq 0 \text{ only for finitely many } n\right\}, \label{eq:c00}
\end{equation}
also called the space of eventually zero sequences. The dual of the space can
now be determined by finding all possible infinitely-dimensional vectors that
have a finite inner product with all elements of $c_{00}$. Since an arbitrary element of
$c_{00}$ has $c_{n > N} = 0$ for some $N$, it has a well-defined inner
product with any infinite-dimensional vector (with finite coefficients), i.e., the
dual space $\Phi^\times$ is simply the space of all sequences without any particular
convergence properties
\begin{equation}
    \Phi^\times = \mathbb{C}^\infty := \left\{\sum_{n = 0}^\infty c_n \ket{n}; \ c_n \in \mathbb{C}, \abs{c_n} < \infty \right\}.
\end{equation}
In particular, we now see that the RP eigenvectors from
Eqs.~\eqref{eq:shift_rp_r}, \eqref{eq:shift_rp_l} are elements of the dual
space\footnote{We chose $\Phi = c_{00}$ because it is the simplest possible choice for which $\R, \L \in \Phi^\times$. It is not the only possible one, nor is it the largest set possible.
Another option would be the set of faster-than-exponentially decaying sequences
$\Lambda_\infty := \left\{\sum_{n = 0}^\infty c_n \ket{n}; \
\sum_{n = 0}^\infty \abs{c_n}^2 r^{2n} < \infty \text{ for all } r >
0\right\} \supset c_{00}$. See Ref.~\cite{meiseIntroFunctional1997}, Chapter 29, for a
treatment of such spaces. Note that the discrete analog of Schwartz
functions~\cite{gelfandGeneralizedFunctionsApplications1964}, the space of
faster-than-polynomially decaying sequences $S := \left\{\sum_{n = 0}^\infty c_n
\ket{n}; \ \mathrm{sup}\, \abs{c_n} n^k < \infty \text{ for all }
k \in \mathbb{N}\right\}$, is not appropriate, since $\R, \L \notin S^\times$.};
$\R, \L \in \Phi^\times$.

Since we identified the appropriate structure for working with $\L$ and $\R$, we
can now precisely determine when the RP description in
Eq.~\eqref{eq:shift_rp_exp} is valid. Since $\R \in \Phi^\times$ (and similar
for $\L$), the inner product $\braket{\psi}{\R}$ is only defined for $\psi \in
\Phi$. In words, the description in Eq.~\eqref{eq:shift_rp_exp} is only valid on
the space of eventually zero sequences $\Phi$. In the coarse-grained many-body
interpretation, these correspond to local observables, meaning that we have
obtained an effective dissipative description of a unitary system on the space
of local observables. 

% In the following section, we shall derive the same result
% directly from the spectral theorem in the rigged Hilbert space.

\subsubsection{Gelfand-Maurin spectral theorem}
\label{sec:shift_spectral}

We now turn to the spectrum of $U$. It is easy to show that $U$ has a purely
continuous spectrum on the entire unit circle $\crc$. As already stated in the
Introduction, applying the standard spectral
theorem~\cite{reedMathPhysI} would, therefore, give us an expression for $U$ via
projection-valued measures. These can be cumbersome to work with, so in this
work, we turn to the equivalent of the spectral theorem in rigged Hilbert
spaces, the so called Gelfand-Maurin spectral
theorem~\cite{gelfandGeneralizedFunctionsApplications1964,
bohmDiracKetsGamow1969}. 

We can check that our choice of $\Phi$ and the operator $U$~\eqref{eq:shift_imp} satisfy the
assumptions of the theorem (more on that later), which then guarantees us that
there exists an analog of the usual (finite-dimensional) spectral decomposition
valid on $\Phi$. More precisely, we have
\begin{equation}
    \mel{\psi}{U^t}{\phi} = \oint_\crc \lambda^t \braket{\psi}{\F_\lambda}\braket{\F_\lambda}{\varphi} \dd\mu(\lambda), \qquad \forall \psi, \phi \in \Phi, \label{eq:shift_nst}
\end{equation}
where $\ket{\F_\lambda} \in \Phi^\times$ for $\lambda \in \sigma(U) = \crc$ are
generalized eigenvectors,
\begin{equation}
    U \ket{\F_\lambda} = \lambda \ket{\F_\lambda}, \label{eq:shift_gen_ev_eq}
\end{equation}
and $\dd\mu(\lambda)$ is a measure on $\crc$ (that must be determined in each
case separately). The generalized eigenvectors also resolve the
identity on $\Phi$, i.e.,
\begin{equation}
    \mel{\psi}{\1}{\phi} = \oint_\crc \braket{\psi}{\F_\lambda}\braket{\F_\lambda}{\phi} \dd\mu(\lambda), \qquad \forall \psi, \phi \in \Phi, \label{eq:shift_id}
\end{equation}
meaning that they can be interpreted as a ``generalized basis'' on $\Phi$.

The generalized eigenvectors can be obtained exactly by solving
Eq.~\eqref{eq:shift_gen_ev_eq} (we derive them for an arbitrary CMV matrix in
Sec.~\ref{sec:cmv_truncation})
\begin{equation}
    % \ket{\F_\lambda} = \ket{0} + \frac{1}{\sin \phi} \left[ \sum_{n = 2, 4, \dots}^\infty \frac{1 - \lambda \cos \phi}{\lambda^{n/2}} \ket{n} + \sum_{n = 1, 3, \dots}^\infty \lambda^{(n - 1)/2} (\lambda - \cos \phi) \ket{n} \right], \label{eq:shift_gen_ev}
    \ket{\F_\lambda}= \ket{0} + \frac{1}{\sin \varphi} \left[\sum_{n = 2, 4, \dots}^\infty \lambda^{n/2 - 1} (\lambda - \cos \varphi) \ket{n} + \sum_{n= 1, 3, \dots}^\infty \frac{1 - \lambda \cos \varphi}{\lambda^{(n + 1)/2}} \ket{n} \right],  \label{eq:shift_gen_ev}
\end{equation}
where we choose the normalization $\braket{0}{\F_\lambda} = 1$. 
% The result is precisely equal to the $\ket{\R(z)}$ from the resolvent~\eqref{eq:shift_res_r}.
We can solve for the measure by satisfying every component of Eq.~\eqref{eq:shift_id}
obtaining\footnote{The obtained measure is nothing but the Verblunsky measure of
the corresponding OPUC, see Sec.~\ref{sec:opuc}}
\begin{equation}
    \dd\mu(\lambda) = \frac{\sin^2 \varphi}{\abs{\lambda - \cos \varphi}^2} \frac{\dd{\lambda}}{2\pi \ii \lambda}.
\end{equation}

We now turn to the generalized spectral decomposition~\eqref{eq:shift_nst}.
While the expression is more convenient than working with projection-valued measures,
the RP resonances are still not immediately evident, but are hidden in the analytic
properties of the generalized eigenvectors. To extract them, we repeat a procedure
similar to the one in Sec.~\ref{sec:shift_res}.
Namely, since we analytically know the integrand, we can solve the integral by the
residue theorem. We do this separately for each component
\begin{equation}
    \mel{\psi}{U^t}{\varphi} = \sum_{n, m = 0}^\infty \braket{\psi}{n} \braket{\phi}{m} \oint_\crc \lambda^t c_n(\lambda) c_m(\lambda)^* \dd\mu(\lambda),
\end{equation}
where we wrote $\ket{\F_\lambda} = \sum_n c_n(\lambda) \ket{n}$.
In order to solve this, we must analytically continue the integrand from the
unit circle to the entire unit disk. The only non-analytic function in the present form is complex
conjugation, which can be analytically continued from the unit circle to the entire complex plane
with the standard trick $z^* \xrightarrow[\text{AC}]{} 1/z$. Applying this
in $c_m(z)^*$ and in the measure $\dd \mu(\lambda)$, we obtain
\begin{equation}
    \mel{\psi}{U^t}{\varphi} = \sum_{n, m = 0}^\infty \braket{\psi}{n} \braket{\phi}{m} \oint_\crc \lambda^t c_n(\lambda) c_m\left(1/\lambda\right) \frac{\sin^2 \varphi}{(\lambda - \cos \varphi)(1/\lambda - \cos \varphi)} \frac{\dd{\lambda}}{2\pi \ii \lambda}.
\end{equation}
By inspecting Eq.~\eqref{eq:shift_gen_ev}, we see that the integrand can diverge
only in $z = \cos \varphi$ and $z = 0$. By the residue theorem
\begin{align}
    \mel{\psi}{U^t}{\varphi} &= \sum_{n, m = 0}^\infty \braket{\psi}{n} \braket{\phi}{m} \mathop{\text{Res}}\limits_{\lambda = \cos \varphi}\left[\lambda^t c_n(\lambda) c_m\left(1/\lambda\right) \frac{\sin^2 \varphi}{\lambda (\lambda - \cos \varphi)(1/\lambda - \cos \varphi)} \right] + \xi \nonumber\\
    &= (\cos \varphi)^t \bra{\psi} \left[ \sum_{n = 0} ^\infty c_n(\cos \varphi) \ket{n} \sum_{m = 0}^\infty c_m(1/\cos \varphi) \bra{m}\right] \ket{\phi} + \xi \nonumber \\
    &= (\cos \varphi)^t \braket{\psi}{\R} \braket{\L}{\phi} + \xi,
\end{align}
where $\xi$ denotes the contributions from the pole at $z = 0$ and $\R, \L$
are defined in Eqs.~\eqref{eq:shift_rp_r}, \eqref{eq:shift_rp_l}. We have thus
reproduced the effective non-unitary description of $U$ on $\Phi$ we obtained
already in Sec.~\ref{sec:shift_res}.

A comment about the assumptions of the Gelfand-Maurin spectral theorem
(sometimes also called the Nuclear spectral theorem) is in order. Firstly, the
theorem assumes that $\Phi$ is a nuclear space. In short, a nuclear space is an
infinite-dimensional generalization of vector spaces distinct from Hilbert
spaces that retains some of the topological properties of finite-dimensional
spaces that Hilbert spaces do not. We will not discuss the
formal aspects in greater detail, for that we refer the reader to
Refs.~\cite{bohmDiracKetsGamow1969, gelfandGeneralizedFunctionsApplications1964}
and to Ref.~\cite{delamadrid_phd} for a construction of the appropriate topology on $c_{00}$ that we
are considering. Importantly, the theorem assumes that $U$ is valid operator
also just on $\Phi$, i.e., $U \Phi \subset \Phi$\footnote{More formally, the
theorem assumes that $U$ is a bounded normal operator on $\mathcal H$, that is
also a continuous operator on $\Phi$, with respect to its (nuclear) topology.}.
This holds in our case, because $U$ acts only locally. For a general unitary
$U$, however, this is a crucial requirement that restricts the choice of the
rigged Hilbert space in which the generalized eigenvectors can be described.
Note that the Gamow vectors do not necessarily lie in the same space as
generalized eigenvectors. In particular, our generalized eigenvectors
$\F_\lambda$, $\lambda \in \crc$ lie in the dual of the space of
faster-than-polynomially decaying sequences $S^\times$ and the choice of $\Phi =
S$ would be a valid choice for the Gelfand-Maurin spectral theorem, whereas the
Gamow vectors $\R, \L$ do not lie in $S^\times$. In all cases that authors are
familiar with, Gamow vectors (i.e., RP eigenvectors) diverge exponentially, see
Ref.~\cite{duhCascade} for intuitive arguments for that in the context of
many-body systems.

\subsection{Non-unitary truncation}
\label{sec:shift_tp}

While the derivation of RP resonances through analytic continuation of the
resolvent or analytic continuation of generalized eigenvectors was possible in
the shift with an impurity~\eqref{eq:shift_imp}, both approaches are in general
hard, especially for chaotic many-body systems. In this section, we consider an
approach that is numerically feasible: diagonalization of a finite version of
$U$. In the present case, (right) eigenvectors of such $U$ can again be
determined exactly. Provided that the finite version of $U$ is equal to the infinite $U$ in all its
rows and columns but the last (i.e., we cut out the top left block of infinite
$U$ and perhaps change the last row/column), they are precisely equal to the
finite truncation of the generalized eigenvectors $\ket{\F_\lambda}$,
Eq.~\eqref{eq:shift_gen_ev}.

Additionally (see Sec.~\ref{sec:cmv_truncation}), we get an equation from the
last row/column, that we can interpret as the boundary condition of the
truncation. We often consider boundary conditions that are unitary, i.e., finite
unitary versions of $U$. In that case, the eigenvalues lie on the unit circle
and, while the information about the decay rate is hidden in the structure of
the eigenvectors (in the same way as in Sec.~\ref{sec:shift_spectral}), it is
hard to extract in practice.

A non-unitary truncation can prove to be more useful for that. We now
consider a truncation that merely projects $U$ to the first $N$ components.
More specifically, let $P_N := \sum_{n = 0}^{N - 1} \ket{n}\bra{n}$ be the projector to
the first $N$ components in $\ell_2$. As already mentioned in Sec.~\ref{sec:cmv_intro}
and explicitly shown in Sec.~\ref{sec:cmv_truncation}, the
eigenvalues of $U_N := P_N U P_N$ with $N \geq 1$ satisfy
\begin{equation}
    \Phi_{N}(\lambda) = \lambda^{N - 1} (\lambda - \cos \varphi) = 0.
\end{equation}
That is, we get a $(N-1)$-times degenerate eigenvalue $\lambda = 0$ and a once
degenerate $\lambda = \cos \varphi$ eigenvalue at the location of the RP
resonance. The corresponding right eigenvector is $\ket{\F_{\lambda = \cos
\varphi}} = \ket{\R}$ and the corresponding left eigenvector is $\bra{\L}$.

Obtaining the RP resonance as the leading eigenvalue of the truncation $U_N$ for
$N \to \infty$ and the Gamow vectors as its leading eigenvectors is not a
fine-tuned phenomenon occurring only in the considered model. In
Sec.~\ref{sec:cmvexp} we show that it is true for a wide range of CMV matrices (but
not all!). As already mentioned in the Introduction, it is also the essence of
the so-called truncated propagator approach of studying quantum
circuits~\cite{prosenRuelleResonancesQuantum2002}. In Ref.~\cite{duhCascade}, we
provided a physical intuition on why the procedure works, but given the results
of the previous sections, a more mathematical intuition is now clear. Truncation
$U_N$ is essentially an effective non-unitary description of dynamics, that is
equal to $U$ up to site $N$. Sending $N \to \infty$ is essentially asking for an
effective non-unitary description for any local dynamics, which is precisely the
RP description obtained in Sec.~\ref{sec:shift_res} and
Sec.~\ref{sec:shift_spectral}.

\section{CMV matrices and orthogonal polynomials on the unit circle}
\label{sec:opuc}

Let us review some facts about CMV matrices and the associated orthogonal polynomials on the unit circle (OPUC)\footnote{Most of the time we will not refer to the original literature; all results can be found in Ref.~\cite{simonOrthogonalPolynomialsUnit2005}, see also
  Refs.~\cite{golinskii,simon_cmv5,golinskiitotik} for shorter overviews.}. At first reading, those not interested in this fascinating field of mathematics can skip to Section~\ref{sec:cmvexp} that explains the physics -- decay of correlation functions -- within a class of CMV matrices with exponentially decaying $\alpha_k$.

CMV matrices are in general infinitely large, however, if one chooses a given
$\alpha_{n-1}$ on a unit circle, $|\alpha_{n-1}|=1$, this will split $U$ into a
direct sum (because $\rho_{n-1}=0$) of a finite CMV unitary of size $n \times n$
(upper left block) and the rest (which is again an infinite CMV matrix).
Alternatively, one can get the same finite unitary CMV matrix by taking a finite
number of $\alpha_k$ and then  ending a finite $\cmvM$ by a $\1_{1\times1}$ block instead of the next $\Theta_k$ (for even $n$). We will be interested in infinite $U$, however, for numerical purposes we will sometimes also use a sufficiently large finite unitary CMV matrix obtained as described above. Given a unitary operator $U$ that is not in a CMV form, it can be written in the CMV form in the basis obtained by Gram-Schmidt orthogonalization of vectors $\{x_0,Ux_0,U^{-1}x_0,U^2x_0,U^{-2}x_0,\ldots \}$, where $x_0$ is a cyclic vector.

CMV matrices are closely connected to OPUC. One way of defining OPUC is via
$\alpha_k$ that determine recurrence relations that orthogonal polynomials
$\Phi_k(z)$ of degree $k$ satisfy,
\begin{equation}
  \alpha^*_{k-1} \Phi_{k+1}(z)=(\alpha^*_k+z \alpha^*_{k-1})\Phi_k(z)-\alpha^*_k \rho^2_{k-1} \Phi_k(z)
\end{equation}
and $\Phi_0(z) := 1$.
$\Phi_k(z)$ have the leading coefficient $1$, i.e., $\Phi_k(z)=z^k+\cdots$ (they
are called monic polynomials). Normalizing $\Phi_n(z)$, one gets orthonormal
polynomials $\varphi_n(z)=\kappa_n \Phi_n(z)$, where $\kappa_n=1/||\Phi_n||$
with $||\Phi_n||^2=\prod_{k=0}^{n-1} \rho_k^2$ (remember
$\rho_k^2=1-|\alpha_k|^2$, Eq.~\eqref{eq:cmv}).

Closely related are the so-called reverse polynomials, also called \S duals,
$\Phi^\star_k(z)$, where the star operation is defined for a polynomial of
degree $n$ by $p^\star_n(z):=z^n p^*_n(1/z^*)$ (beware of the difference between
the two stars: on the LHS the star $\star$ denotes the \S dual, on the RHS the
``multiplication'' star $*$ is the complex conjugation!). More explicitly, if
$p_n(z)=\sum_{j=0}^n c_j z^j$, then $p^\star_n(z)=\sum_{j=0}^n c^*_j z^{n-j}$
(coefficients are ``reversed''). Written for a pair $\Phi_k(z)$ and
$\Phi^\star_k(z)$, or for $\varphi_k(z)$ and $\varphi^\star_k(z)$, the recursion
is even simpler, namely, what is called the \S recursion reads
\begin{equation}
    \begin{pmatrix}
        \varphi_{k+1}(z) \\
        \varphi^\star_{k+1}(z)
    \end{pmatrix} =\mathrm{A}_k(z) \begin{pmatrix}
      \varphi_k(z)\\
      \varphi^\star_k(z)
    \end{pmatrix},\qquad \mathrm{A}_k(z)=\frac{1}{\rho_k}\begin{pmatrix}
    z & -\alpha^*_k \\
    -z\alpha_k  & 1
    \end{pmatrix}.
    \label{eq:opuc_iter}
\end{equation}
The above recursion can be iterated and polynomials expressed in terms of the transfer matrix~\cite{golinskiinevai} $T_n(z)$ as
\begin{equation}
    \begin{pmatrix}
        \varphi_{n}(z) \\
        \varphi^\star_{n}(z)
    \end{pmatrix} =\mathrm{T}_{n}(z) \mathbf{e}_+,\quad \mathbf{e}_+:=\begin{pmatrix} 1 \\
    1
    \end{pmatrix},\quad \mathrm{T}_n(z):=\mathrm{A}_{n-1}(z)\cdots \mathrm{A}_0(z).
\end{equation}
Occasionally, one also needs polynomials of the second kind, denoted by $\Psi_k(z)$ for the monic ones and $\psi_k(z)$ for the normalized ones. $\Psi_n(z)$ are obtained for a measure in which all Verblunsky coefficients have an opposite sign, that is, if $\Phi_n(z)$ are for $\{ \alpha_k \}$, then the associated $\Psi_n(z)$ are for $\{ -\alpha_k \}$. They can be written as
\begin{equation}
    \begin{pmatrix}
        \psi_{n}(z) \\
        -\psi^\star_{n}(z)
    \end{pmatrix} = \mathrm{T}_{n}(z) \mathbf{e}_{-},\quad \mathbf{e}_{-}:=\begin{pmatrix} 1 \\
    -1
    \end{pmatrix}. \label{eq:psi_def}
\end{equation}
A variation of transfer matrices will be useful in
Section~\ref{sec:transfer}. The main part in recursion relations is
multiplication by $z$, and a CMV matrix is nothing but a representation of this
multiplication in the basis of orthogonal polynomials. Explicitly, a matrix
element of a CMV matrix is
\begin{equation}
    U_{k,p}=\ev{\chi_k, z \chi_p}_\mu = \ev{\xopuc_p, z \xopuc_k}_\mu,
\end{equation}
where
\begin{align}
  \chi_{2k}(z) &:=z^{-k}\varphi^\star_{2k}(z), \qquad \qquad \chi_{2k-1}(z) :=z^{-k+1}\varphi_{2k-1}(z), \label{eq:chi_def} \\
  \xopuc_{2k}(z) &:= z^{-k} \varphi_{2k}(z), \qquad \qquad \xopuc_{2k - 1}(z) := z^{-k} \varphi_{2k - 1}^\star(z) \label{eq:x_def}
\end{align}
make up the so-called CMV basis and the alternate\footnote{Another common notation
for the alternate basis is $\xopuc_k(z) \equiv x_k(z)$.} CMV basis, respectively.
CMV matrices are therefore an explicit embodiment of the spectral
theorem~\cite{reedMathPhysI} that says that every unitary with a cyclic
vector\footref{ft:cyclicity} on a Hilbert space is unitarily equivalent to a
multiplication by $z$, $f(z) \to zf(z)$, on the unit circle under a measure
$\dd\mu$, i.e., space $L^2(\mathbbm{T},\dd\mu)$.

Another useful object is the \S function $D(z)$ defined as
\begin{equation}
  D(z) = \lim_{n \to \infty} \frac{1}{\varphi^\star_n(z)}.
  \label{eq:Dz}
\end{equation}
For our choices of $\alpha_k$, for which the measure\footnote{One should not
confuse the measure $\dd\mu$ with the multiplicative factor $\mu$ in exponential
Verblunsky coefficients $\alpha_k$ (\ref{eq:exp}).} $\dd\mu$ under which
$\Phi_n(z)$ are orthogonal is continuous, we can simply express it
as\footnote{Alternatively one can also use the Carath\' eodory function $F(z)$,
$w(\theta)=\frac{1}{2\pi} {\rm Re}\left[F(\e^{\ii \theta})\right]$.}
\begin{equation}
  \dd\mu(\theta)=w(\theta)d\theta,\qquad w(\theta)=\frac{1}{2\pi} |D(\e^{\ii \theta})|^2,
\end{equation}
with normalization $\int_0^{2\pi}w(\theta)\dd\theta=1$. Polynomials $\varphi_n$
are orthonormal with respect to this measure, $\int_0^{2\pi} \varphi_n^*(\e^{\ii
\theta}) \varphi_k(\e^{\ii \theta})w(\theta)\dd\theta=\delta_{k,n}$. As already
written in Eq.~\eqref{eq:Cmu}, the correlation function in the 1st basis vector
$C(t)=\langle x_0 | U^t | x_0\rangle = [U^t]_{0,0}$ is equal to a $t$-th moment
of the measure, $C(t)=\int_0^{2\pi} \e^{-\ii t \theta}w(\theta)\dd\theta$.
Writing the Carath\' eodory function $F(z)$, which is a particular transformation
of the measure $\dd\mu$, one has
\begin{equation}
  F(z):=\int \frac{\e^{\ii \theta}+z}{\e^{\ii \theta}-z} \dd\mu(\theta)=1+2\left[ C(1) z+C(2) z^2+\cdots \right],
  \label{eq:Fz}
\end{equation}
where the last equality is obtained by Taylor expanding the argument of the
integral. The Carath\' eodory function can also be expressed in terms of
orthogonal polynomials as
\begin{equation}
  F(z)=\lim_{n \to \infty} \frac{\Psi^\star_n(z)}{\Phi^\star_n(z)}.
  \label{eq:FzP}
\end{equation}
% where $\Psi_n(z)$ are polynomials of the second kind. 
The above
Eq.~\eqref{eq:FzP} has an important guarantee that the convergence is uniform and
that a finite-$n$ approximation has an error of order $z^{n+1}$. Such a
guarantee is not available for $D(z)$ in Eq.~\eqref{eq:Dz}. 

Looking at Eq.~\eqref{eq:Fz}, one could infer the asymptotic decay of $C(t)$ by
studying Taylor series of $F(z)$ at large $z$, or start directly from the
measure and its moments. For our purposes, though, this is not the best way. While
there is a one-to-one correspondence between $\{ \alpha_k\}$ and the
measure\footnote{Given a measure $\dd\mu$ one can determine the corresponding
OPUC. A neat way is variational: $\Phi_n(z)$ are polynomials $p_n(z)$ of order
$n$ that minimize $\int_{|z|=1} |p_n(z)|^2 \dd\mu(z)$.} $\dd\mu$, as well as
with the zeroes of OPUC\footnote{Given a set of $n$ points $z_j \in
\mathbbm{D}$, there exists a measure $\dd\mu$ such that the corresponding
orthogonal polynomial has exactly those $z_j$ for zeroes, $\Phi_n(z_j)=0$.},
there are better ways.

One can connect $\Phi_n(z)$ with $C(t)$ in two different ways. A direct one proceeds by defining a Toeplitz matrix $C^{(n)}_{j,k}$ given by correlations, $C^{(n)}_{j,k}:=C(j-k)$, $0 \le j,k \le n-1$ ($C(t)$ for the cyclic vector $x_0$), and its determinant $D_n:=\det(C^{(n+1)}_{j,k})$. Then one has
\begin{equation}
  \Phi_n(z)=\frac{1}{D_{n-1}}\det(C^{(n+1)}(z)),
\end{equation}
where the matrix $C^{(n+1)}(z)$ is equal to $C^{(n+1)}$ except that the last-row elements in $C^{(n+1)}$, which are $C(n),C(n-1),\ldots,C(0)$, are replaced by $1,z,\ldots,z^n$. Normalization of $\Phi_n$ can be also written as $\kappa_n^2=D_{n-1}/D_n$, and $|\alpha_n|^2=1-\frac{D_{n-1}D_{n+1}}{D_n^2}$.

The second way, most useful for our truncated propagator method, is by
truncating an infinite $U$ to a finite leading (upper-left) block of size $N$,
obtaining $U_N$. $U_N$ is also called a cutoff CMV. Such finite $U_N$ -- a
truncated CMV of size $N \times N$ obtained by projecting $U$ to a subspace -- is
subunitary\footnote{And is different than taking a finite CMV by setting
$\alpha_{N-1}=1$, in which case $U$ is still unitary.} with all its eigenvalues
being within the unit disk. The orthogonal polynomial $\Phi_N(z)$ is exactly
equal~\cite{simon_cmv5} to the characteristic polynomial of $U_N$,
\begin{equation}
  \Phi_N(z)=\det(z-U_N).
  \label{eq:detU}
\end{equation}
Therefore, the eigenvalues of the truncated $U_N$ are equal to the zeros of
$\Phi_N(z)$. Denoting those zeroes by $z_j \in \mathbbm{D}$ such that
$\Phi_N(z)=\prod_{j=1}^N (z-z_j)$, the reverse polynomial $\Phi^\star_N(z)$ has
zeroes at $1/z^*_j$. Zeroes of OPUC have been studied a lot, and they are
important also for RP resonances. 

Namely, provided the truncation size $N$ is
large enough so that finite-size effects are avoided, $C(t)$ evaluated at finite
$t$ with $U_N$ instead of with $U$ in Eq.~\eqref{eq:Cx} will be unchanged.
Therefore, having in mind a spectral decomposition of a finite $U_N$, like in
Eq.~\eqref{eq:U}, the RP resonances should be given by the zeroes of $\Phi_N(z)$, specifically, the leading resonance $\lambda$ should be equal to $z_1$, the largest (in modulus) zero
of $\Phi_N(z)$ (in the limit $N \to \infty$). While this might usually be the
case, we identify a phase (phase II in Fig.~\ref{fig:presek}) where the above
argument fails. The reason is that $U_N$ is not normal and in such cases one can
not always expect that its eigenvalues will converge to the spectrum of an infinite $U$ as $N\to \infty$, nor that the eigenvalues of
$U_N$ (zeroes of $\Phi_N(z)$), no matter how large an $N$, are a
good indication of the properties of $U_N$, e.g., of the asymptotic decay rate of $C(t)$. If the matrix is very badly
conditioned, i.e., its eigenvalues are highly sensitive to small perturbations, it might be better to look at the pseudospetrum~\cite{trefethen}. We shall in fact see that the situation in phase II is in some ways rather similar to a recently studied
instance~\cite{boombust} where it is the pseudospectrum and not spectrum that matters. However, in our case the situation is more nuanced. The pseudospectrum, which is a robust generalization of the spectrum, is agnostic to locality of states, whereas the locality of observables is absolutely crucial for $C(t)$. Looking just at the pseudospectrum will not be enough -- the asymptotic decay $|\lambda|$ will not be equal to the largest modulus in the pseudospectrum (which in our case approaches 1 as $N\to\infty$) --  and one has to find another way. Fortunately, there is a theorem (Theorem 7.1.2 in Ref.~\cite{simonOrthogonalPolynomialsUnit2005}) that provides just that. Omitting details, it states that
\begin{equation}
  \limsup_{n \to \infty} |C(t)|^{1/n}=1/R_F=1/R_D,
  \label{eq:thm}
\end{equation}
where $R_F$ is the convergence radius of the Carath\' eodory function $F(z)$
around $z=0$ and $R_D$ is the convergence radius of the \S function $D(z)$ around
$z=0$ ($F(z)$ and $D(z)$ are both analytic in $\mathbbm{D}$ so
$R_F,R_D>1$). Considering that the LHS of Eq.~(\ref{eq:thm}) is by definition equal to $|\lambda|$, the theorem therefore says that $|\lambda|$ is equal to the reciprocal convergence radius of either $D(z)$ or $F(z)$. However, one tricky point remains.

Looking at Eq.~\eqref{eq:Dz} and Eq.~\eqref{eq:Fz}, unless one has exact analytic
expressions, which we do not, one will have to approximate both functions by
finite order polynomials. One might be tempted to conclude that the convergence
radius will be given by the zeroes of $\Phi^\star_n(z)$, i.e., poles in
finite-$n$ $D(z)$, and therefore the convergence radius of both should be
$|1/z_1|$. This, however, we already dismissed as not working in the problematic
phase II. Where is the catch? The finite-$N$ problem of eigenvalues of $U_N$
here re-appears in the way the limiting process works in the definition of both
functions. The Carath\' eodory case is simpler so let us explain it first. As we
mentioned, in Eq.~\eqref{eq:Fz} we know that the error at finite $n$ is of order
$z^{n+1}$. Therefore, we can trust $\Psi^\star_n/\Phi^\star_n$ to be correct
only up to order $z^n$. What one has to do is expand the rational function
$\Psi^\star_n(z)/\Phi^\star_n(z)$ in a Taylor series around $z=0$, putting all
the correct terms up to order $n$ in a finite-$n$ approximation $F_n(z)$
(dropping higher order terms), and then write
\begin{equation}
  F(z) = \lim_{n \to \infty} F_n(z),\qquad \frac{\Psi^\star_n(z)}{\Phi^\star_n(z)}=:F_n(z)+{\cal O}(z^{n+1}).
  \label{eq:Fn}
\end{equation}
Such $F_n(z)$ is a polynomial of order $n$ and will give a correct approximation of $F(z)$ at any finite $n$ in the sense of correctly approximating the convergence radius of $F(z)$.

For $D(z)$, the situation is in principle similar with an added difficulty that we do
not know what is the error of a finite $n$ approximation in Eq.~\eqref{eq:Dz}.
What one can nevertheless do is Taylor expand $1/\varphi^\star_n(z)$ to some
order $m$ smaller than $n$, putting all those terms in $D_{n,m}(z)$ (polynomial
of order $m$), and then write a double limit\footnote{The two limits do not commute.},
\begin{equation}
  D(z)=\lim_{m \to \infty} \lim_{n \to \infty} D_{n,m}(z),\qquad \frac{1}{\varphi^\star_n(z)}=:D_{n,m}(z)+{\cal O}(z^{m+1}).
  \label{eq:Dnm}
\end{equation}
Provided $m$ is chosen so that it is smaller than the error, i.e., that one can
bound $|D(z)-D_{n,m}(z)| < {\cal O}(z^{m+1})$, divergence of a finite-$n$
$D_{n,m}(z)$ will correctly approximate $R_D$ of $D(z)$. Although we do not know
the order of the error, empirically, in our specific case, often some fraction like $m \sim n/2$ works well.

\section{CMV matrices with exponentially decaying coefficients}
\label{sec:cmvexp}

Let us start by illustrating rich relaxation and the behavior of the leading RP
resonance one can get with simple CMV matrices. A given CMV matrix $U$
(\ref{eq:cmv}) is specified by Verblunsky coefficients $\alpha_k \in
\mathbbm{D}$, where $\mathbbm{D}=\{ z\, ; |z|<1\}$ is an open unit disk. Such
$U$ is an infinite-dimensional unitary on $\ell_2$. We want to pick the simplest
possible choice of $\alpha_k$ in order to, hopefully, be able to get exact
results, but at the same time still have a realistic description of potentially
real systems.

In our context of relaxation, where $U$ is a unitary (super)propagator on the
space of operators, one must think of basis elements as being operators. We have
seen in Section~\ref{sec:shift} that a single nonzero $\alpha_0$ results in
operator dynamics that is similar to what happens in dual-unitary circuits,
where, at a sufficiently large operator support $r$, the structure of generalized
(Gamow) eigenvectors (eigenvectors of the truncated $U$) is exactly that of a
shift. In generic chaotic systems one still has an isolated leading RP
resonance, but the shift property for eigenvectors holds only asymptotically as
$r \to \infty$~\cite{duhCascade}. With that in mind we are going to take
$\alpha_k$ that decay exponentially with index $k$,
\begin{equation}
  \alpha_k = a\cdot \mu^{k}, \qquad k = 0, 1, \dots
  \label{eq:exp}
\end{equation}
As in Sec.~\ref{sec:shift}, in order to connect with the physics of lattice
systems, one should think of the basis as a coarse-grained description such that
the $2n$-th and $(2n+1)$-th basis elements represent all operators with support
on $r=n$ consecutive sites. As we shall see, the amplitude $a$ will strongly influence the decay rate of correlations, while one can think of $\mu$ as determining the operator support length scale on which dynamics
becomes shift-like (or, equivalently, the length scales on which the stable and
unstable manifold effectively separate). 

The main object we want to study is the decay of correlation functions in some vector (operator) $x$,
\begin{equation}
  C(t)=\langle x | U^t | x\rangle \asymp \lambda^t,
  \label{eq:Cx}
\end{equation}
and in particular its asymptotic exponential decay rate $\lambda$ that should be equal to the leading RP resonance (we will call $\lambda$, the asymptotic decay of $C(t)$, the leading RP resonance). For CMV matrices (\ref{eq:cmv}) the cyclic vector is simply the first basis vector $\ket{x_0} = \ket{0}$, representing operators with support on a single site. Our default choice of $x$ will be $x_0$, meaning that we are calculating autocorrelation function of single-site operators. Another possibility is to take the $k$-th basis vector $\ket{x_k} = \ket{k}$ (representing operators with larger support). \diff{Note that in Ref.~\cite{yehPRB25}, the connection between exponentially decaying correlation functions and exponentially decaying Verblunsky coefficients has been studied, however, always in our phase I.}

\subsection{The phase diagram}
\label{sec:phases}
\begin{figure}[t!]
  \centerline{\includegraphics[width=0.9\linewidth]{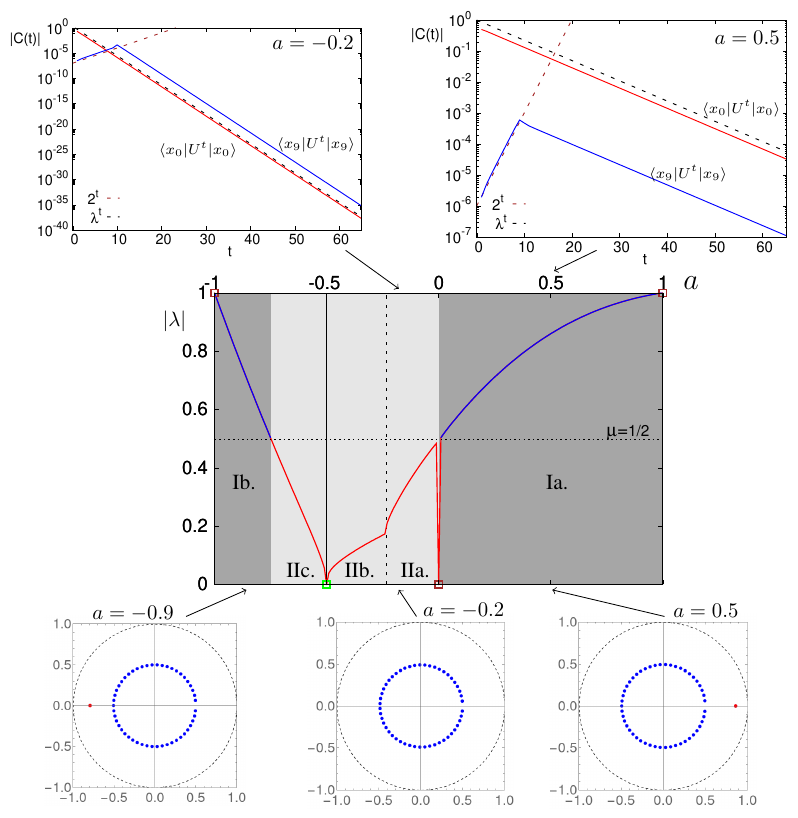}}
  \caption{Autocorrelation functions (\ref{eq:Cx}) (top), its asymptotic decay $\lambda$ (middle), and eigenvalues of the truncated $U_{50}$ (bottom) for CMV matrices with exponential $\alpha_k$ (\ref{eq:exp}) at $\mu=\frac{1}{2}$. Middle: Phases as a function of $a$. In phase I (dark grey) $\lambda$ is equal to the largest eigenvalues of $U_{N}$ (bottom row, the red point), while in phase II (light grey) all eigenvalues of $U_{N}$ are on a circle with radius $\mu=\frac{1}{2}$ and do not give the correct $\lambda<\frac{1}{2}$. Top: the asymptotic decay of $C(t)$ agrees with $\lambda$ read off from the middle plot (dashed black lines). Details of subphases Ia and Ib, and IIa,b,c, will be discussed in Section~\ref{sec:phases}.}
    \label{fig:presek}
\end{figure}
In Fig.~\ref{fig:presek}, top row, we show the autocorrelation
function\footnote{Calculated numerically by iterating a sufficiently large CMV
$U$ so that there are no finite-size effects.} for two choices of $x$, seeing
that, in both cases, the long-time decay is nicely exponential. For $x_k$ with
some fixed $k$, $C(t)$ initially grows as $(1/\mu)^t$ until $t=k$, when it
starts to decay as $\lambda^t$. The long-time asymptotic decay is, therefore, for initial states (operators) localized below some finite $k$, always exponential with the same rate.

In the middle plot, we show the dependence of this $\lambda$ calculated numerically. We do not calculate it by fitting exponential function to $C(t)$, but rather by using orthogonal polynomials\footnote{For specific values of $a$ highlighted in the figure one obtains $\lambda\approx 0.860, 0.263, -0.794$ at $a=0.5, -0.2, -0.9$, respectively.} associated to the CMV matrix, specifically the Carath\' eodory function $F(z)$. What we do is we calculate a finite-size approximation $F_n(z)$ according to Eq.~\eqref{eq:Fn}, from which we then numerically extract the convergence radius\footnote{Specifically, we use Jentzsch theorem~\cite{jentzsch} connecting zeroes of partial sums with a convergence radius (see also Appendix~\ref{app:Dnm}) and set $|\lambda|=\langle |\frac{1}{z_j}|\rangle$, where the average is over the bulk of zeroes $z_j$ of $F_{300}(z)$.} $R_F$ giving us $|\lambda|$ according to the Theorem in Eq.~\eqref{eq:thm}. In general, we observe a continuous dependence of the leading RP resonance $|\lambda|$ on $a$, though with a non-continuous 1st derivative and some special points highlighted by brown and green squares. Perhaps the most interesting one the green square at $a=-\mu$ (also vertical black line), at which $\lambda=0$ and $C(t)$ exhibits a Gaussian instead of an exponential decay. By a vertical
dashed line at $a \approx -0.235$ we also highlight a point of discontinuity in
the derivative $\dd\lambda/\dd a$. We can in fact distinguish two phases, the phase I and the phase II, which can be furthermore split into subphases Ia and Ib, and IIa, IIb and IIc. We postpone a detailed discussion of all
differences between the phases to subsequent subsections, for now let us highlight the main and the most interesting difference, namely, in the spectrum of the truncated propagator $U_N$.

In the bottom row of Fig.~\ref{fig:presek} we show all eigenvalues of a
truncated CMV matrix $U_{N}$ obtained by taking a finite $N \times N$ upper-left
block of an infinite $U$. One would expect that the largest eigenvalue
$\lambda_1$ (shown as a red point) of $U_N$ should be equal to the asymptotic
decay $\lambda$. This is indeed the case in phase I, but not in phase II. As we
can see in the bottom middle figure, in phase II, all eigenvalues $\lambda_j$ of $U_N$
are in fact on a circle with radius $\mu$, whereas the correct RP resonance
$|\lambda|$ is strictly smaller than $\mu$ in all of phase II (middle frame in
Fig.~\ref{fig:presek}). As
we will see, the crux lies in the ring of eigenvalues (blue), all of which are
very badly conditioned, and which acts as an ``event horizon'', hiding the true
resonance inside the ring. Some analytical tricks are required to lure $\lambda$
out of its hiding place. 

The overall structure of the eigenvalues of $U_N$, seen in the bottom row, can
in fact be related to rigorous theorems. We can, for instance, see that in all
phases (and for all $\mu$ and $a$, not just for the shown examples), all but at
most one eigenvalue of the truncated CMV matrix $U_N$ for large $N$, i.e., zeroes of
$\Phi_N(z)$, condense on a circle of radius equal to $\mu$.
Defining $b:=\lim_{n \to \infty} \sup |\alpha_n|^{1/n}$, which for our
exponential case is just $b=\mu$, a theorem
says~\cite{saff90,nevaitotik,simon04I} (see also Chapter 8 and 7.1 in
Ref.~\cite{simonOrthogonalPolynomialsUnit2005}) that in the limit $N \to \infty$
a measure $1$ of zeroes of $\Phi_N(z)$ uniformly converges to a circle of radius
$b$, which we are going to call the ring, and that there is at most a finite
number of zeroes $z_j$ that are larger, $|z_j|>b$. Moreover, as observed
numerically (bottom Fig.~\ref{fig:presek}), the zeros on the ring are clock-like
uniformly spread~\cite{simon04I}. We note that while for our exponential
$\alpha_k$ there are no zeros inside the ring, in general there can be. Namely,
if $\alpha_k$ are sums of exponentials, one can have zeros also inside the ring,
see Ref.~\cite{simon04II}. The ``outlier'' zeroes $|z_j|>b$ are called the
Nevai-Totik points, have a well defined limit as $N \to \infty$, and are nothing
but the RP resonances. For our specific exact exponential form
of $\alpha_k$, we numerically observe that there is at most one such Nevai-Totik
point, specifically, exactly one in phase I, in which the asymptotic decay of
$C(t)$ is slower than that of $\alpha_k$, and zero in phase II where we have an
opposite situation with $C(t)$ decaying faster than $\alpha_k$, i.e.,
$|\lambda|<\mu$.

\begin{figure}[t!]
  \centerline{\includegraphics[width=0.5\linewidth]{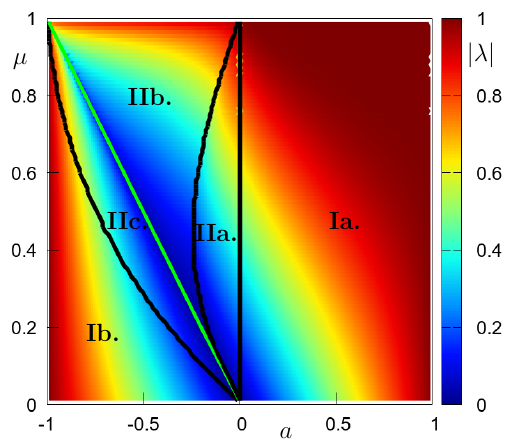}}
  \caption{The phase diagram for exponential $\alpha_k=a\cdot \mu^{k}$. The asymptotic decay $\lambda$ of autocorrelations (\ref{eq:Cx}) in $x_0$, i.e., the leading RP resonance, is shown as a function of $\mu$ and $a$. Green line is a special phase with a Gaussian decay, whereas black curves delimit different phases with exponential decay.}
  \label{fig:fazni}
  % exp_matrika_n300.dat: n=300, razvoj do 300
\end{figure}

In Fig.~\ref{fig:fazni}, we show the phase diagram for all $a$ and $\mu$ (in
Fig.~\ref{fig:presek} we showed a cross-section at $\mu=\frac{1}{2}$). In the
following we are going to discuss details of the phases.

\subsubsection{Phase I}

The main characteristic feature of phase I is that the largest eigenvalue $\lambda_1$ of the truncated $U_N$ lies outside the ring (is larger than $\mu$) and correctly predicts the asymptotic decay of $C(t)$. That is, one has
\begin{equation}
  \lambda=\lambda_1.
\end{equation}
The leading RP resonance is therefore equal to the largest Nevai-Totik point
$z_1$, i.e., the largest zero of $\Phi_N(z)$. One can make a finer
classification. In the phase denoted by Ia in Fig.~\ref{fig:presek}, one has
$\lambda>0$ and $C(t)$ for $x_0$ is always positive, whereas, in the phase Ib,
$\lambda$ is negative, $\lambda<0$ (see bottom left Fig.~\ref{fig:presek}), and
$C(t)$ changes sign between even and odd times (for $x_9$ and $t\le 9$ in the
initial growth phase, $C(t)$ is negative in both Ia and Ib).

There are also some special points at the edges of phase I. At $a=\pm 1$, the CMV
matrix $U$ trivially decouples into a direct sum of a $1\times 1$ identity block
and the rest that is again a CMV matrix. If $a=0$, causing all $\alpha_k=0$, one
gets a free CMV matrix~\cite{simonOrthogonalPolynomialsUnit2005} which has
$w(\theta)=\frac{1}{2\pi}$, the orthogonal polynomials are $\Phi_n(z)=z^n$ and
$\Phi^\star_n(z)=1$, and therefore $D(z)=F(z)=1$, making correlation function in
the cyclic $x_0$ decay immediately, $C(t)=\delta_{t,0}$. This is reflected in a
discontinuous jump in $\lambda$ at $a=0$ in Fig.~\ref{fig:presek}.

While we do not have a simple closed form expression for the dependence of $\lambda$ on $a$ and $\mu$, one can get an approximation by explicitly calculating $C(t)$ at small $t$ and approximating $|\lambda| \approx |C(t+1)/C(t)|$. For instance, one gets $C(1)=a$, $C(2)=a^2+a(1-a^2)\mu$, $C(3)=a^3+a(1-a^2)[2a\mu+(1-a^2)\mu^2-a^2\mu^4]$. Such an approximation works well for not too large $\mu$, and the better the larger $t$ one uses, for instance, approximating $|\lambda| \approx |C(3)/C(2)|$ the absolute error in $|\lambda|$ in the whole phase I is less than $0.01$ for $\mu<0.5$.

While the truncated propagator $U_N$ gives the correct leading RP resonance, it
is instructive to look at how one would get it via the convergence radius $R_D$
of the \S function $D(z)$ according to the Theorem in Eq.(\ref{eq:thm}). In
phase I, things are simple: simply approximating $D(z) \approx D_{n,\infty}(z)$,
where $D_{n,\infty}(z):=1/\varphi^\star_n(z)$ (\ref{eq:Dz}), the convergence
radius of $D(z)$ will be given by the pole closest to the origin, i.e., the
smallest zero of $\varphi^\star_n(z)$. Because the zeroes of
$\varphi^\star_n(z)$ are equal to $1/z_j^*$, where $z_j$ are zeroes of
$\varphi_n(z)$, we see that one has a ring of poles at radius $1/\mu$ (due to
the ring of zeroes) as well as the relevant nearest pole at distance
$1/|z_1|=R_D<\mu$. If one would instead use a finite-$n$ approximation
$D_{n,m}(z)$ (\ref{eq:Dnm}), the ring of poles at $1/\mu$ immediately moves due
to their high sensitiviy (large condition number) inside to the radius $R_D$,
being visible as zeroes of $D_{n,n}(z)$. See Appendix~\ref{app:Dnm} for details
and figures.

The method of expanding $D_{n,\infty}(z)$ into a power series to order $m<n$, thereby obtaining $D_{n,m}(z)$, will be crucial to get the correct leading RP resonance in the phase II.

\subsubsection{Phase II}

In phase II, there are no isolated eigenvalues of the truncated CMV $U_N$, i.e.,
zeroes of $\Phi_n(z)$, only the ring of eigenvalues at radius $\mu$, despite the
decay still being exponential with a continuous dependence of $\lambda$ on $a$
(Fig.~\ref{fig:presek}). Considering that $\abs{\lambda}<\mu$ throughout phase II,
it looks like the RP resonance is ``hidden'' inside the ring. To get this hidden
$\lambda$ we use different procedures: analytic continuation will be
discussed in Section~\ref{sec:transfer}, while here we focus on using the \S or
the Carath\' eodory function and the theorem in Eq.~\eqref{eq:thm}.

As opposed to phase I, simply approximating $D(z) \approx D_{n,\infty}(z)$ does not work in phase II because the convergence radius of $D_{n,\infty}(z)$ is at $1/\mu$ due to all zeroes of $\varphi_n(z)$ being at the ring with radius $\mu$. Here one really has to expand $D_{n,\infty}(z)$ to some smaller order $m$ (what $m$ one should take seems to depend on $a$ and $\mu$). The convergence radius of such $D_{n,m}(z)$ then does give the correct convergence radius $R_D$ and $|\lambda|$, see Appendix~\ref{app:Dnm} for details. Because $m$ is a priori not known, an even better method is expansion of a finite-$n$ approximation of the Carath\' eodory function $F(z)$. For $F(z)$ we namely have a guarantee (\ref{eq:Fn}) that expanding to order $m=n$ will work and give the correct convergence radius and $|\lambda|$.

There is also a very interesting special point in phase II at $a=\mu$ where $\lambda=0$ and the decay of autocorrelation function for the cyclic vector $x_0$ is Gaussian,
\begin{equation}
  C(t)=(-1)^t \mu^{t^2}.
\end{equation}
For the initial vector $x_k$ one would initially have an exponential growth
$\sim (1/\mu)^t$, and only after $t=k$ the Gaussian decay $\sim \mu^{(t-k)^2}$.
Note that, in terms of the eigenvalues of $U_N$, nothing special happens at this
point -- all eigenvalues of $U_N$ are still equidistantly distributed around the
ring at $\mu$. The orthogonal polynomials at this special point
$\alpha_k=-\mu^{k+1}$ are known as the Rogers-\S
polynomials~\cite{simonOrthogonalPolynomialsUnit2005}, in fact they are the
rotated Rogers-\S polynomials as the original ones have $\alpha_k=(-1)^k
\mu^{k+1}$. Namely, if one rotates the measure, that is makes a transformation
$\dd\mu \to \sigma \dd\mu$ with $|\sigma|=1$, the Verblunsky coefficients are
transformed as $\alpha_k \to \sigma^{-(k+1)}\alpha_k$. For instance, taking
$\sigma=\e^{\ii \pi}$ multiplies $\alpha_k$ by $(-1)^{k+1}$. A rotated measure
also rotates the zeroes of $\Phi_n(z)$ and so that would be the way to rotate
the leading RP resonance out of the real axis.

In fact, it turns out that in phase IIb, the leading resonance is not on the real
axis anymore. This can be seen from the signs that $C(t)$ takes (here discussed
only for $x_0$). Namely, in phase IIa, $C(t)$ is negative for any $t$, in phase
IIc, $C(t)$ alternates in sign between even and odd times due to a negative
$\lambda$, whereas in phase IIb, $C(t)$, which is real, oscillates with a
nontrivial period that continuously varies as a function of $a$ at fixed $\mu$.
This is due to a complex pair of leading resonances $|\lambda|\e^{\pm \ii
\beta}$ (one needs a complex pair to get real $C(t)$), see also Fig.~\ref{fig:rpr_diagram}.

\subsection{Finite-size effects}
\label{sec:finite_size}

\begin{figure}[t!]
  \centerline{\includegraphics[width=0.47\linewidth]{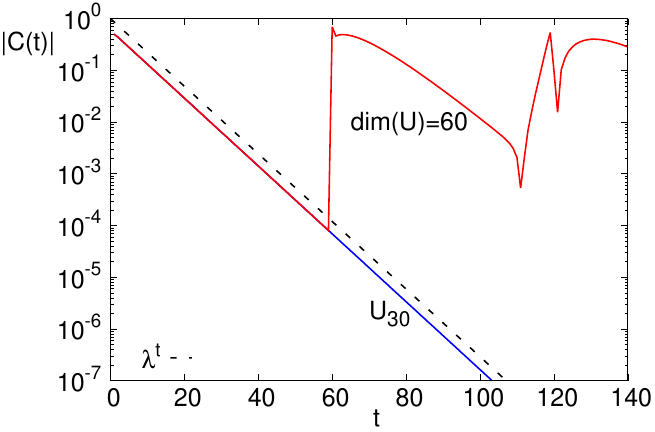}\hskip4mm\includegraphics[width=0.47\linewidth]{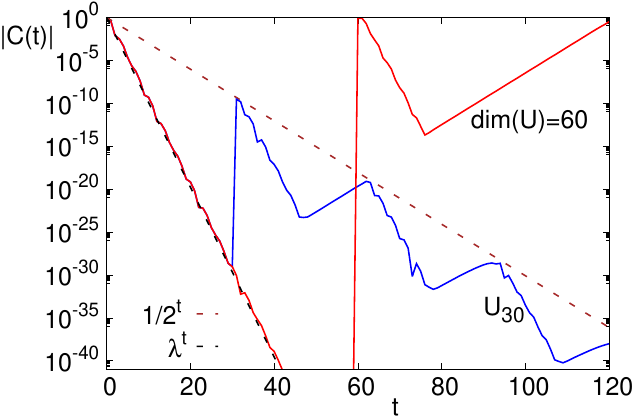}}
  \caption{Finite-size effects in phase I (left, $a=\frac{1}{2}$) and phase II (right, $a=-\frac{2}{5}$), in both cases for $\mu=\frac{1}{2}$. Autocorrelation function for $x_0$ (\ref{eq:Cx}) is shown for a finite unitary CMV of size $60$ (red/top curves), and for a truncated $U_{N=30}$ (blue/bottom curves). In phase II one has a boom-bust-like behavior for the nonunitary truncated $U_{30}$: until $t=N$ the decay is the correct asymptotic $\lambda^t$ ($\lambda \approx 0.1026$), whereas later on finite-size effects cause the slower average decay given by the ring at $\mu$.}
    \label{fig:finite}
\end{figure}

Let us finally discuss finite-size effects, that is, how does $C(t)$ look
like if one calculates it either with a finite unitary CMV, or a finite
truncated $U_N$, instead of with an infinite $U$. This is relevant from a
practical point of view -- if one would study $C(t)$ in a given physical system
where one can only numerically study truncated propagators with modest operator
supports (remember that in CMV matrices the size $N$ should be thought of as the
maximal operator support), finite-size effects would be important. As
we will now see, finite-size effects also give us an independent insight into
differences between phases I and II. Autocorrelation $C(t)$ for a finite unitary
CMV, as well as for a (non-unitary) truncated $U_N$, are shown in
Fig.~\ref{fig:finite}.
Using a finite unitary $U$ of size $N$, finite-size effects appear in $C(t)$ at
$t>N$ (60 in the Figure), after which $C(t)$ exhibits revivals, in line with the
unitarity of $U$. Even more interesting is the situation with the truncated $U_N$. In
phase I, the autocorrelation functions decays exponentially even for $t>N$, i.e.,
there are no revivals because $U_N$ is not unitary (see bottom
Fig.~\ref{fig:presek}). In the phase II, though, at $t=N$ (and its multiples) one
gets a partial revival so that the average decay of those revivals follows a
slower decay $\sim \mu^t$ determined by the ring of eigenvalues. In the
incorrect thermodynamic limit of $\lim_{N \to \infty}\lim_{t \to \infty}$, the
decay of $C(t)$ is given by the largest eigenvalue of $U_N$ (the ring), whereas
in the correct thermodynamic limit of $\lim_{t \to \infty}\lim_{N \to \infty}$,
it is given by the leading hidden RP resonance $\lambda$. This is the same as
what has been called a phantom relaxation found for entanglement evolution in
some random circuits~\cite{phantom,staircase}. Finite jumps observed in
Fig.~\ref{fig:finite}, right, are analogous to a similar behavior in Figs.~13-14 of
Ref.~\cite{phantom}, and can be traced back to a ring of badly conditioned
eigenvalues~\cite{boombust}. While in such situations it is sometimes the
pseudospectrum rather than the spectrum that determines the correct relaxation
rate~\cite{boombust}, in our phase II, this does not seem to be the case. The
pseudospectrum of our truncated CMV matrix $U_N$ is namely in the limit $N \to
\infty$ equal to the unit disk, both in phase I and phase II. It remains an open
problem to better understand mathematical consequences that such a ring can
have, and also why are its effects so different between phases I and II.
Understanding more intuitively why and under which conditions such rings of
eigenvalues occur is also open\footnote{One situation when a ring appears is
perturbing a $n\times n$ Jordan block by a perturbation of strength $\epsilon$
under which an $n$-times degenerate eigenvalue splits into $n$ eigenvalues on a
ring of radius $\epsilon^{1/n}$ -- the Puiseux series~\cite{kato}.}.

\subsection{Transfer matrix approach}
\label{sec:transfer}

In previous sections, we obtained the leading RP resonance of CMV matrices via
theorems from the OPUC literature. In this section, we study the CMV matrices
using a transfer matrix approach similar to the one in
Refs.~\cite{golinskiinevai, gesztesyWeylTitchmarsh2006, weikardInverseResonanceProblem2010}. The
central results are eigenvectors of the truncated propagator in
Sec.~\ref{sec:cmv_truncation} and an exact transfer matrix expression for the
resolvent we derive in Sec.~\ref{sec:cmv_resolvent}. Through it, we will once
again obtain the leading RP resonance, as well as subleading RP resonances
(Sec.~\ref{sec:cmv_ac}) and Gamow vectors (Sec.~\ref{sec:cmv_gamow}). Note that
an exact expression for the resolvent for $\abs{z} < 1$ in terms of OPUC was
known before~\cite{simonOrthogonalPolynomialsUnit2005}. We will rederive this
and obtain the expression for $\abs{z} > 1$. Additionally, the form we obtain is
more convenient to study the analytic continuation of the resolvent and Gamow
vectors.

\subsubsection{Eigenvectors of (finite) CMV matrices}
\label{sec:cmv_truncation}

Let us first study the eigenvalues and eigenvectors of CMV matrices. In
particular, let's consider the truncated propagator. We
are solving the equation
\begin{equation}
    U_N \ket{\Fr_\lambda} = \lambda \ket{\Fr_\lambda}. \label{eq:cmv_ev_eq}
\end{equation}
Due to $U = \cmvL \cmvM$ being a product of two block diagonal matrices, we can
approach the problem using an auxiliary degree of freedom
\begin{equation}
    \ket{v} := \cmvM \ket{u}, \label{eq:cmv_ev_v_eq}
\end{equation}
which we can understand as a ``half-step'' of the entire $U$. Writing out our
vectors by components $\ket{\Fr_\lambda} = \sum_{n=0}^N u_n \ket{n}$ and
$\ket{v} = \sum_{n = 0}^N v_n \ket{n}$, Eqs.~\eqref{eq:cmv_ev_eq} and
\eqref{eq:cmv_ev_v_eq} can be compactly written using transfer matrices
\begin{equation}
    \begin{pmatrix}
        u_{n}\\
        v_{n}
    \end{pmatrix} = \T_{n-1}(\lambda) \begin{pmatrix}
        u_{n-1}\\
        v_{n-1}
    \end{pmatrix},  \quad 0 < n < N, \hspace{17pt}
    \T_k(z) := \frac{1}{\rho_k} \begin{cases}
        \begin{pmatrix}
            - \alpha_k & 1/z \\
            z & -\alpha_k^*
        \end{pmatrix}; & k \text{ even} \\
        \begin{pmatrix}
            - \alpha_k^* & 1 \\
            1 & -\alpha_k
        \end{pmatrix}; & k \text{ odd}
    \end{cases} \label{eq:cmv_tm}
\end{equation}
with boundary conditions
\begin{align}
    &u_0 = v_0  \label{eq:cmv_ev_bc} \\
    v_{N - 1} = \alpha_{N-1}^* u_{N - 1}, \quad \text{ if } N \text{ is even}&, \qquad \qquad 
    \alpha_{N-1}^* v_{N - 1} = z u_{N - 1}, \quad \text{ if } N \text{ is odd}. \nonumber
\end{align}
For the derivation see Appendix~\ref{app:cmv_ev}.

By further defining a product of transfer matrices
\begin{equation}
    \M_k(z) := \T_{k-1}(z) \dots \T_1(z) \T_0(z) \label{eq:cmv_mm}
\end{equation}
and $\M_0(z) := \1_{2 \times 2}$, the system of equations~\eqref{eq:cmv_tm},
\eqref{eq:cmv_ev_bc} can be used to derive explicit results for the right
eigenvectors $\ket{\Fr_\lambda}$ and (using an analogous procedure) left
eigenvectors $\bra{\Fl_\lambda}$,
\begin{align}
    \ket{\Fr_\lambda} &= \sum_{n = 0}^{N - 1} \ve_1 \M_n(\lambda) \vep \ket{n} = \sum_{n = 0}^{N - 1} \xopuc_n(\lambda) \ket{n}, \label{eq:cmv_evr}\\
    \bra{\Fl_\lambda} &= \sum_{n = 0}^{N - 1} (-1)^n \vem \M^{-1}_n(\lambda) \ve_1 \bra{n} = \sum_{n = 0}^{N - 1} \chi_n(\lambda) \bra{n}, \label{eq:cmv_evl}
\end{align}
where we defined the basis vectors $\ve_1 := \begin{pmatrix}
    1 \\ 0
\end{pmatrix}$, $\ve_2 := \begin{pmatrix}
    0 \\ 1
\end{pmatrix}$, $\ve_\pm := \ve_1 \pm \ve_2$\footnote{For
clarity, we omit the transposition in products between vectors and matrices,
i.e., we write $\ve_i \M \ve_j \equiv \ve_i^\top \M \ve_j$.} and the OPUC $\chi_k, \xopuc_k$ are defined in Eqs.~\eqref{eq:chi_def},
\eqref{eq:x_def}. The relations between transfer matrix expressions
and OPUC are derived in Appendix~\ref{app:cmv_opuc}. The eigenvalues are given as
solutions to the following equation
\begin{equation}
    \boldsymbol v_\text{BC} \M_{N}(\lambda) \vep = \lambda^{\lfloor N/2 \rfloor} \varphi_N(\lambda) = 0, \label{eq:cmv_tm_ev} \qquad \qquad \text{where }
    \boldsymbol v_\text{BC} = \begin{cases}
        \ve_1; & N \text{ even} \\
        \ve_2; & N \text{ odd}
    \end{cases}.
\end{equation}
The OPUC expression is a known result we have already stated in Sec.~\ref{sec:cmv_intro}.

Provided that we are in phase I, Eqs.~\eqref{eq:cmv_evr}, \eqref{eq:cmv_evl} can be used to obtain
the leading Gamow vectors $\ket{\R} = \ket{\Fr_{\lambda_1}}$, $\bra{\L} =
\bra{\Fl_{\lambda_1}}$, where $\lambda_1$ is the leading RP resonance. In the
following, we shall see that the expression is valid also in phase II (if
we set $\lambda$ to be the leading RP resonance) and study the properties
of Gamow vectors in Sec.~\ref{sec:cmv_gamow}.

Eigenvector expression~\eqref{eq:cmv_evr} is valid also in the infinite case if
we send $N \to \infty$, i.e., the generalized eigenvectors of a CMV matrix
are $\ket{\F_\lambda} = \ket{R_\lambda}, \lambda \in \crc$. One can use them in
the Gelfand-Maurin spectral theorem and analytically continue them to obtain the
RP resonances and Gamow vectors, similar as we did for shift with an impurity in
Sec.~\ref{sec:shift_spectral}. If we normalize
$\braket{0}{\F_\lambda} = 1$, the appropriate measure one must use in
the Gelfand-Maurin spectral theorem is precisely the Verblunsky measure (see
Sec.~\ref{sec:opuc}). We
will not pursue this approach in the present paper.

\subsubsection{Exact calculation of the resolvent}
\label{sec:cmv_resolvent}

In order to better understand phase II, we now turn to the resolvent formalism
introduced in Sec.~\ref{sec:shift_res}. To get the resolvent of the infinite CMV
matrix $\mel{\phi'}{R(z)}{\phi} = \mel{\phi'}{(z - U)^{-1}}{\phi}$, we solve for
$\ket{u} := (z - U)^{-1} \ket{\phi}$, i.e.,
\begin{equation}
    U\ket{u} = z \ket{u} - \ket{\phi}.
\end{equation}
This is an inhomogeneous version of the eigenequation we solved in
Sec.~\ref{sec:cmv_truncation}. We explicitly solve it in
Appendix~\ref{app:cmv_ev} (defining a half-step auxiliary degree of freedom
$\ket{v}$ in the same way as in Eq.~\eqref{eq:cmv_ev_v_eq}), obtaining
\begin{equation}
    \begin{pmatrix}
        u_{n} \\ v_{n}
    \end{pmatrix} = \M_{n}(z) \vep v_0 - \frac{1}{z}\M_{n}(z) \sum_{k = 0}^{2\lfloor n/2\rfloor-1} (-1)^k \M^{-1}_{k}(z) \ve_1 \phi_{k}, \label{eq:cmv_iter_res}
\end{equation}
where we denoted $\ket{u} = \sum_n u_n \ket{n}$ and similarly for $\ket{v}$ and $
\ket{\phi}$. $\M_n(z)$ is defined in Eq.~\eqref{eq:cmv_mm}.

Our expression still contains a free parameter $v_0$, which is essentially a
normalization. In the case of the non-unitary truncation in
Sec.~\ref{sec:cmv_truncation}, it was set by the boundary
condition~\eqref{eq:cmv_ev_bc} at the site of the truncation. Here, we determine
it by fine-tuning it in a way that make $R(z)$ bounded, yielding a similar
``boundary condition at infinity''. Namely, since $R(z)$ is an analytic function
outside the spectrum, we required that there $u_{n},v_n \to 0$ as $n\to \infty$.
The behavior of $u_n, v_n$ in the limit depends on the parity of $n$, here we
consider $n = 2N$ to always be even\footnote{This is a choice, we will later
send $N \to \infty$. One could just as well take $n = 2N + 1$ which would lead
to slightly different boundary conditions ($\T_n(z)$ for large $n$ exchanges $u$
and $v$), but a physically equivalent result.}. As $n\to\infty$, we have
$\alpha_n\to 0$ and thus for large $N$
\begin{equation}
    \T_{2N-1}(z)\T_{2N-2}(z) \to \begin{pmatrix}
        z & 0\\
        0 & 1/z
    \end{pmatrix}. \label{eq:t_asymp}
\end{equation}
Let us first consider $\abs{z} > 1$, which we will use for analytic
continuation. In this case, $u_{2N}$ will blow up as $\sim z^N$ unless it is
fine-tuned to vanish, yielding the ``boundary condition at infinity'' $u_{2N\to
\infty} =0$. For $\abs{z} < 1$, we instead have $v_{2N}$ diverging as $\sim
z^{-N}$, and we have to enforce $v_{2N \to \infty} = 0$. We apply these boundary
conditions to Eq.~\eqref{eq:cmv_iter_res} in Appendix~\ref{app:cmv_resolvent},
obtaining the explicit expression for the resolvent
\begin{align}
    R(z) &= \ket{\R(z)} f(z) \bra{\L(z)} + \Gamma(z), \label{eq:cmv_resolvent} \\
    f(z) &:= \lim_{N \to \infty} \begin{cases}
       \frac{1}{2z}\frac{\ve_1 \M_{2N}(z)\vem}{\ve_1 \M_{2N}(z)\vep}; & \abs{z} > 1 \\
       \frac{1}{2z}\frac{\ve_2 \M_{2N}(z)\vem}{\ve_2 \M_{2N}(z)\vep}; & \abs{z} < 1
    \end{cases} = \lim_{N\to \infty}\begin{cases}
        \frac{1}{2z} \frac{\psi_{2N}(z)}{\varphi_{2N}(z)}; \abs{z} > 1 \\
        \frac{1}{2z} \frac{\psi^\star_{2N}(z)}{\varphi^\star_{2N}(z)}; \abs{z} < 1
    \end{cases}, \\
    \ket{\R(z)} &:= \sum_{n = 0}^\infty \ve_1 \M_n(z) \vep \ket{n} = \sum_{n = 0}^\infty \xopuc_n(z) \ket{n}, \\
    \bra{\L(z)} &:= \sum_{m = 0}^\infty (-1)^m \vem \M^{-1}_m(z) \ve_1 \bra{m} = \sum_{n = 0}^\infty \chi_n(z) \bra{n}.
\end{align}
where $\Gamma(z)$ is a matrix valued function with no
poles for $z > 0$, its exact expression is written in Eq.~\eqref{eq:cmv_zeta}.
The expressions with OPUC were obtained using the results in Appendix~\ref{app:cmv_opuc}.

The $f(z)$ function inside the unit circle can be expressed with the
Carath\' eodory function $F(z)$ (see Sec.~\ref{sec:opuc}), $f(z; \abs{z} < 1) =
\frac{1}{2z} F(z)$, compatible with the exact expression for the resolvent
inside the unit circle in Ref.~\cite{simonOrthogonalPolynomialsUnit2005}.
Provided that OPUC for our choice of $\alpha_k$ are well-behaved, we can also
express the $f(z)$ function outside the unit circle with $F(z)$, $f(z; \abs{z} >
1) = \frac{1}{2z} F(1/z^*)^*$. Recalling Sec.~\ref{sec:shift_res}, the poles of
$f_2(z)$, the analytic continuation of $f$ to the second Riemann sheet (i.e.,
continuing from $\abs{z} > 1$ to $\abs{z} \leq 1$), will be the RP resonances
$\lambda$. The fact that the poles of $f_2(z)$ and $f_1(z)$ (the analytic
continuation of $f$ to the first Riemann sheet, from $\abs{z} < 1$ to $\abs{z}
\geq 1$) are related by a simple $z \mapsto 1/z^*$ transformation is intuitively
expected. Namely, one can derive an analog to the resolvent
identity~\eqref{eq:res_id} for negative powers $U^{-t} = -\frac{1}{2 \pi \ii}
\oint_{\mathcal C'} z^{-t} R(z) \dd{z}$, where $\mathcal C'$ is a curve just \textit{inside} the
unit circle. By an analogous procedure to the one in Sec.~\ref{sec:shift_res},
one then obtains the result that the RP resonances for negative times are
the inverse values of poles of $f_1(z)$. Since for unitary matrices, reversing the
time amounts to simply taking the Hermitian adjoint, one expects that the
absolute values of forward-time and backward-time RP resonances are the same,
i.e., the poles of $f_1$ and $f_2$ are related by transforming $z \mapsto 1/z^*$.
This also gives us an understanding of theorem~\eqref{eq:thm}, which relates
the leading RP resonance to the convergence radius of $F(z)$ expanded around $z = 0$,
$\lambda = 1/R_F$. $1/ R_F$ in the present case is simply the largest pole of $f_1(z)$
(lying outside the unit circle). In the following, we shall do the analytic continuation
more carefully, obtaining also the Gamow vectors and subleading RP resonances, which
goes beyond theorem~\eqref{eq:thm}.

\subsubsection{Analytic continuation of the resolvent}
\label{sec:cmv_ac}

We now turn to analytically continuing the expression for the resolvent in
Eq.~\eqref{eq:cmv_resolvent}. It is easy to see that the only part of the
expression which can potentially have non-trivial poles is $f(z)$, since all
other expressions contain only polynomials in powers of $z$ and $1/z$. As in
Sec.~\ref{sec:shift_res}, we continue it from $\abs{z} > 1$ to $\abs{z} \leq 1$.
Naively, we just use the expression for $\abs{z} > 1$, obtaining the
analytic continuation on the second Riemann sheet
\begin{equation}
    \tilde f_2(z) = \lim_{N\to\infty} \frac{1}{2z}\frac{\ve_1 \M_{2N}(z)\vem}{\ve_1 \M_{2N}(z)\vep}. \label{eq:cmv_f_ac_naive}
\end{equation}
Using the results from Sec.~\ref{sec:shift_res}, the RP resonances
$\lambda_i$ are the poles of $f_2(z)$ and the corresponding
Gamow vectors are
\begin{align}
    \ket{\R_i} &= \ket{\R(\lambda_i)} = \sum_{n = 0}^\infty \ve_1 \M_n(\lambda_i) \vep \ket{n}, \label{eq:cmv_gamow_r} \\
    \bra{\L_i} &= \bra{\L(\lambda_i)} = \sum_{m = 0}^\infty (-1)^m \vem \M^{-1}_m(\lambda_i) \ve_1 \bra{m}, \label{eq:cmv_gamow_l}
\end{align}
Both results exactly coincide with the truncated propagator results in Eqs.
\eqref{eq:cmv_evr}, \eqref{eq:cmv_evl}, \eqref{eq:cmv_tm_ev} (remember, in our
$N \to \infty$ limit we keep $N$ even), meaning that the resolvent formalism
gives the correct leading RP resonance in phase I.

We numerically observe that Eq.~\eqref{eq:cmv_f_ac_naive} converges only for
$\abs{z} > \mu$, meaning that it is the correct analytic continuation
only for $\abs{z} > \mu$. Note that the
expression in Eq.~\eqref{eq:cmv_f_ac_naive} is a ratio of two power series that
can have a finite radius of convergence. Namely, in the first non-trivial order
in $\alpha_k$ the numerator and denominator of $\tilde f_2(z)$ are
\begin{equation}
    \ve_1 \M_{2N \to \infty}(z) \ve_\pm \approx z^N \left[c_1^{(\pm)} - c_2^{(\pm)} \sum_{k = 0}^{2N-1} \alpha_k^* \tfrac{1}{z^k}\right] = z^N \left[c_1^{(\pm)} - c_2^{(\pm)} a\sum_{k = 0}^{2N-1} \tfrac{{\mu}^{k}}{z^k}\right], \label{eq:cmv_res_poles_intuition}
\end{equation}
where $c_{1, 2}^{(\pm)}$ are constant determined by $\alpha_k$ at small $k$ that
do not affect the asymptotics. See Appendix~\ref{app:cmv_asymp} for the
derivation and details. The expression in Eq.~\eqref{eq:cmv_res_poles_intuition}
converges for $\abs{z} > \mu$ (the $z^N$ prefactors from the numerator and
denominator cancel out). However, $\tilde f_2(z)$ does not diverge as $\abs{z}$
approaches $\mu$ from above, and we can analytically continue further.

\begin{figure}[ht]
    \centering
    \includegraphics[width=0.45\linewidth]{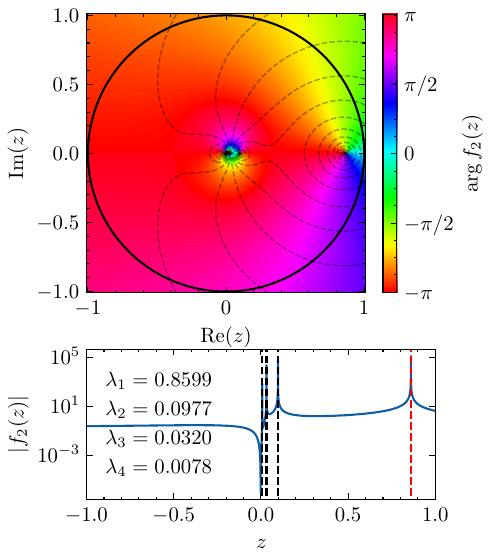}
    \hspace{10pt}
    \includegraphics[width=0.45\linewidth]{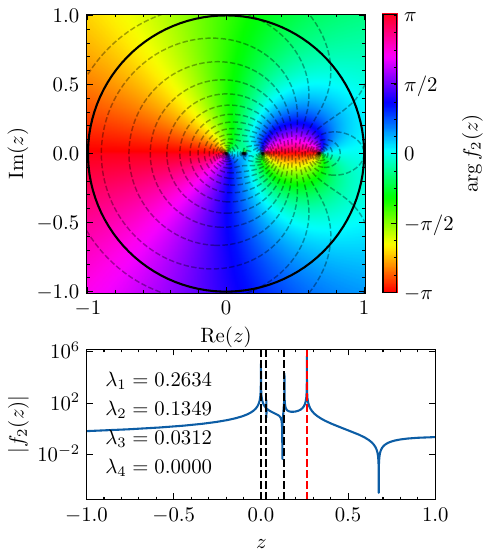}
    \caption{Numerical analytic continuation on the second Riemann sheet
    $f_2(z)$ of the $f(z)$ function from the resolvent \eqref{eq:cmv_resolvent},
    $a = 0.5, \mu = 0.5$ (left) and $a = -0.2, \mu = 0.5$ (right). Top plots
    show the phase of $f_2(z)$, dashed lines show the contour lines of
    $|f_2(z)|$. Bottom plots show $|f_2(z)|$ on the real axis along with
    the precise location of the poles (dashed lines). Obtained by fitting a rational function
    to $f(z)$ just outside the unit circle, each point there is evaluated by
    truncating to the first $N = 100$ terms in Eq.~\eqref{eq:cmv_f_ac_naive}. }
    \label{fig:cmv_resolvent}
\end{figure}

While we do not have an analytical result for that, we can do it by numerically
fitting a rational function to $f(z)$ just outside the unit circle, where the
power series is converging, and using that expression as the analytical
continuation $f_2(z)$\footnote{We use the adaptive Antoulas-Anderson (AAA) rational
function approximation~\cite{aaa2018, aaa2024} implemented in
\texttt{RationalFunctionApproximation.jl} library~\cite{driscoll2025rational}}.
We can then visualize $f_2(z)$ and accurately extract its poles. The result for
two different CMV matrices, one from phase I and one from phase II, is shown in
Fig.~\ref{fig:cmv_resolvent}. The leading pole accurately reproduces the results
obtained from OPUC theory in Sec.~\ref{sec:phases} in both phases.

\begin{figure}[ht]
    \centering
    \includegraphics[width=0.6\linewidth]{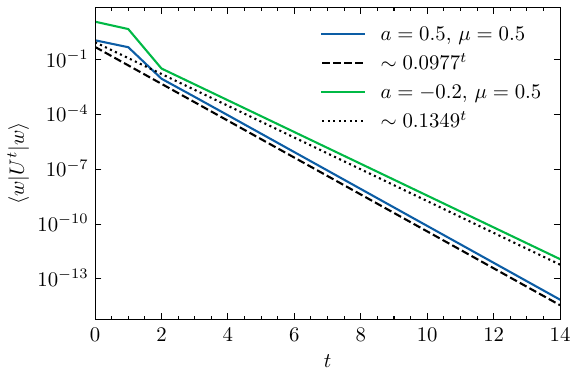}
    \caption{Correlation functions of $\ket{w} = \frac{\alpha_1 -
    \lambda_1}{\rho_1} \ket{0} + \ket{1}$ for two different CMV matrices along
    with the fitted asymptotic decay rates matching the subleading RP resonances
    $\lambda_2$ in Fig.~\ref{fig:cmv_resolvent}.}
    \label{fig:cmv_subleading}
\end{figure}

Additionally, the full analytic continuation of the resolvent gives us access to
subleading poles of the resolvent. To demonstrate that they are actually also
subleading RP resonances, we study the decay rate of correlation functions of vectors
orthogonal to the leading Gamow vector $\ket{\R_1}$. Knowing the exact expression
for the Gamow vector from Eq.~\eqref{eq:cmv_gamow_r}, one of the simplest options
is $\ket{w} = \frac{\alpha_1 - \lambda_1}{\rho_1} \ket{0} + \ket{1}$. Correlation
function in $\ket{w}$ is shown in Fig.~\ref{fig:cmv_subleading} along with the
fitted decay exponents, which match the subleading RP resonance $\lambda_2$
in Fig.~\ref{fig:cmv_resolvent} perfectly.

\subsubsection{Scaling of Gamow vectors}
\label{sec:cmv_gamow}

Using the exact expressions for Gamow vectors in Eq.~\eqref{eq:cmv_gamow_r} (and
already in Eq.~\eqref{eq:cmv_evr} from the truncated propagator) and the
asymptotic expansion of transfer matrices in Appendix~\ref{app:cmv_asymp}, we now
study the asymptotic scaling of Gamow vectors.

Due to the staggered structure of CMV matrices, let us first consider the even
components. In that case, using Eqs.~\eqref{eq:cmv_gamow_r} and
\eqref{eq:cmv_asymp_odd}, we get for large $N$ in first order in $\alpha_k$
\begin{align}
    \abs{\braket{2N + 1}{\R_i}} &= \abs{\ve_1 \M_{2N + 1}(\lambda_i) \vep} \approx \frac{1}{\abs{\lambda_i}^{N + 1}} \left[-c_1^{(+)} a\sum_{k = 0}^{2N} \mu^k \abs{\lambda_i}^k + c_2^{(+)}\right] \nonumber \\
    &\sim \text{const.} \times \abs{\lambda_i}^{-N} + \text{const.} \times \left(\mu^2\abs{\lambda_i}\right)^N \sim \abs{\lambda_i}^{-N}.
\end{align}
where $c_{1, 2}^{(+)}$ are constant determined by $\alpha_k$ at small $k$ that
do not affect the asymptotics (Appendix~\ref{app:cmv_asymp}).
Since we have $\abs{\mu} < 1$ and $\abs{\lambda_i} < 1$ in both phases, the
leading scaling is $\abs{\lambda_i}^{-N}$ and the even components thus diverge.

Similarly, using Eqs.~\eqref{eq:cmv_gamow_r} and \eqref{eq:cmv_asymp_even}, we 
obtain for odd components
\begin{align}
    \abs{\braket{2N}{\R_i}} &= \abs{\ve_1 \M_{2N}(\lambda_i) \vep} \approx 
    c_1^{(+)}\abs{\lambda_i}^{N} - c_2^{(+)} a\abs{\lambda_i}^{N}\sum_{k = 0}^{2N-1} \tfrac{\mu^{k}}{\abs{\lambda_i}^k} \nonumber \\
    &\sim \text{const.} \times \abs{\lambda_i}^{N} + \text{const.} \times (\mu^2/ \abs{\lambda_i})^N,
\end{align}
The leading scaling now depends on the values of $\lambda$ and $\mu$. In phase
II, where $\abs{\lambda} < \mu$, the leading scaling is $\sim
(\mu^2/\lambda)^N$, whereas in phase I with $\abs{\lambda} > \mu$, a naive guess
for leading scaling would be $\sim \lambda_i^N$. At the point of the RP
resonances this is wrong, however, since the expression for the eigenvectors
components is precisely equal to the denominator of the resolvent in
Eq.~\eqref{eq:cmv_res_poles_intuition} and the subleading scaling dominates. In
other words, at the RP resonance,
constants $c_{1, 2}^{(+)}$ must be fine-tuned in such a way that $\ve_1 \M_{2N}(\lambda_i) \vep$ vanishes in $N \to \infty$. The scaling is then given by the deviation
from the $N  \to \infty$ limit, that is
\begin{equation}
    \braket{2N}{\R_i}_\text{phase I} \sim \lambda_i^N \sum_{k = 2N + 2}^\infty \frac{\alpha^*_k}{\lambda^k} \sim \left(\frac{\mu^2}{\lambda_i}\right)^N,
\end{equation}
which is precisely the same as scaling in phase II.

One can repeat a similar procedure for left Gamow vectors to
obtain the final result for both phases
\begin{align}
    \abs{\braket{n}{\R_i}}^2 &\sim \begin{cases}
        \left(\mu^2/\abs{\lambda_i}\right)^n;& n \text{ even} \\
        \abs{\lambda_i}^{-n};& n \text{ odd}
    \end{cases}, \label{eq:cmv_scaling_r}\\
    \abs{\braket{\L_i}{n}}^2 &\sim \begin{cases}
        \abs{\lambda_i}^{-n};& n \text{ even} \\
        \left(\mu^2/\abs{\lambda_i}\right)^n;& n \text{ odd}
    \end{cases}. \label{eq:cmv_scaling_l}
\end{align}
Due to the similarities to many-body systems, we shall refer to matrix elements
$\braket{n}{\R_i}$, $\braket{\L_i}{n}$ as partial norms. Both in CMV matrices
and many-body systems, we observe exponential divergence which encodes that
initially local operators ($\ket{n}$ for small $n$) spread to increasingly
nonlocal ones ($\ket{n}$ for large $n$) -- a chaos scenario for both spin
chains and discrete unitary models, such as the considered CMV model. The
divergence rate is a function of $\lambda_i$, which is a consequence of
unitarity. For a detailed discussion of that see Ref.~\cite{duhCascade}.
The diverging structure of Gamow vectors can be formalized with rigged
Hilbert spaces in the same way as in Sec.~\ref{sec:shift_rigged}. An
appropriate choice is again $\Phi = c_{00}$ and $\R, \L \in \Phi^\times$,
meaning that the RP resonances once again give an effective theory of
local operators (i.e., elements of $\Phi = c_{00}$, see Eq.~\eqref{eq:c00}).

\begin{figure}[ht]
    \centering
    \includegraphics[width=0.48\linewidth]{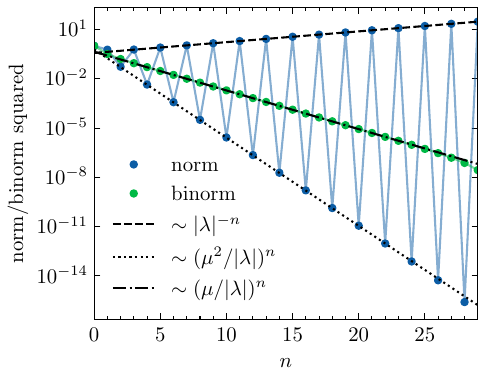}
    \includegraphics[width=0.48\linewidth]{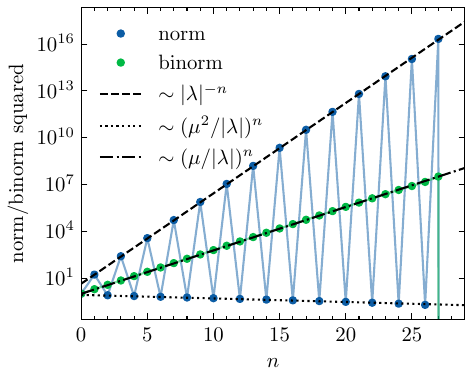}
    \caption{Scaling of partial norms of the right Gamow vector and binorms, $a = 0.5, \mu = 0.5$
    (left) and $a = -0.2, \mu = 0.5$ (right). Phase I (left) was obtained by
    diagonalization of the truncated propagator for $N = 30$ and phase II
    (right) by iterating $\ket{\R_1} \approx U_N^t \ket{0} / \norm{U_N^t
    \ket{0}}$ with $N = 30$ to $t=14$, just before hitting the boundary.}
    \label{fig:cmv_gamow}
\end{figure}

The scalings of partial norms of right Gamow vectors
are numerically tested in both phases in Fig.~\ref{fig:cmv_gamow}, where we
observe good agreement. Gamow vectors can be numerically evaluated in different
ways; in both phases we can use the transfer matrix expression
\eqref{eq:cmv_gamow_r}, whereas in phase I, we can also obtain them as
leading eigenvectors of the truncated propagator. In both phases, one can also obtain the
leading Gamow vector by iterating the dynamics on a vector $\ket{w}$ initially
localized near $n = 0$, i.e., $\ket{\R_1} \approx U^t \ket{w} / \norm{U^t
\ket{w}}$ for large $t$. In phase I, if one takes the truncated propagator, this
is essentially the power method for finding the leading eigenvalue of it, whereas in
phase II one must be careful to iterate only up to times $t$ for which
$U^t\ket{w}$ has not yet reached the boundary (see also Sec.~\ref{sec:finite_size}
for a discussion of finite-size effects). We have verified that all
approaches give the same Gamow vectors.

An interesting quantity to study is also the overlap between left and right
eigenvectors, the so-called partial binorm
\begin{equation}
    \abs{\braket{\L_i}{n}\braket{n}{\R_i}} \sim (\mu/\abs{\lambda_i})^n,
\end{equation}
the scaling of which we have deduced from Eqs.~\eqref{eq:cmv_scaling_r},
\eqref{eq:cmv_scaling_l}. We test the scaling numerically in
Fig.~\ref{fig:cmv_gamow}. The partial binorm of the leading Gamow vector has a physical
interpretation, namely, by writing $\ket{\R_i} \sim U^t \ket{0}$ and $\bra{\L_i}
\sim \bra{0}U^{-t}$ for large $t$, the partial norm can be expressed as
\begin{equation}
    \abs{\braket{\L_i}{n}\braket{n}{\R_i}} \sim \abs{\mel{0}{U^{-t}}{n}\mel{n}{U^t}{0}}.
\end{equation}
In words, the partial binorm is the probability for a ``backflow echo'', i.e.,
for $\ket{0}$ initially localized on the boundary of the infinite system to grow
to some large $\ket{n}$ in time $t$ and then shrink again to $\ket{0}$ in the same time $t$. It is
intuitively clear that when the probabilities of such processes are small (i.e.,
the partial binorms decay with $n$), approximating an infinite system by a
finite system will be good. In CMV matrices, the partial binorms decay in phase I and
diverge in phase II, which is compatible with the fact that the truncated
propagator correctly predicts the leading RP resonance only in phase I.
In more realistic many-body systems, the binorms were found to generically
decay~\cite{duhCascade}, giving an intuitive reason why the truncated propagator
gives correct predictions there. It would be
interesting to find and study a many-body system with diverging binorms of the
leading Gamow vector.

We finally remark on the scaling of the condition numbers of eigenvectors of the
truncated propagator $\kappa_i := \frac{\norm{\mathfrak R_i} \norm{\mathfrak
L_i}}{\abs{\braket{\mathfrak L_i}{\mathfrak R_i}}}$. The condition number is
usually introduced to bound the change of the corresponding eigenvalues under
perturbations, but in Ref.~\cite{duhCascade} we argued that in order for RP
resonances to give an effective theory valid for local operators, $\kappa_i$
must diverge in the same way as partial norms. While in the present paper, we
argued for that through the rigged Hilbert space formulation, it is indeed also
easy to see that $\kappa_i \sim \abs{\lambda_i}^{-N}$ for large enough $N$,
where $N$ is the truncation size, compatible with the divergence of partial
norms~\eqref{eq:cmv_scaling_r}.

\section{Conclusions}

We exactly determined Ruelle-Pollicott (RP) resonances of two families of
infinite unitary CMV matrices describing operator evolution in discrete time. In
both cases we clarified the underlying mathematical structures leading to an
effective dissipative description of unitary dynamics at large times, $U^t
\approx \sum_i \lambda_i^t \ket{\R_i} \bra{\L_i}$, where $\lambda_i$ are RP
resonances and $\R_i$, $\L_i$ are the Gamow vectors (RP eigenvectors), valid for
local operators. Because any unitary operator (with a cyclic vector) takes a CMV
form in an appropriate basis, CMV matrices can be used as a convenient
representation of, in principle, any specific model -- one just has to pick
appropriate Verblunsky coefficients $\alpha_k$. While in the present work we do
not target any concrete model, rather, we study two toy examples, we
nevertheless use Verblunsky coefficients that are motivated by real physics. We
especially focused on the rich physics of the case with exponentially decaying
Verblunsky coefficients\footnote{Exponentially decaying Verblunsky coefficients
are in a way the most interesting ``critical'' case that separates simpler
situations. Namely, in CMV with power-law decaying Verblunsky coefficients, the
rate of decay of correlation functions matches the rate of decay of Verblunsky
coefficients \diff{(proposition 7.1.1 in Ref.}~\cite{simonOrthogonalPolynomialsUnit2005}), while in CMV matrices
with faster-than-exponential decay of Verblunsky coefficients, there is a
one-to-one correspondence between the RP resonances and Verblunsky
coefficients~\cite{weikardInverseResonanceProblem2010}.} $\alpha_k \sim \mu^k$, where we identify two phases.
The results
in phase I match the heuristic derivations and conjectures obtained for generic
many-body systems in Ref.~\cite{duhCascade}. For instance, the rate of decay of
correlation functions and the rate of divergence of both partial norms and
condition numbers (i.e., the ``spatial fractality'' of Gamow vectors) are not
independent -- they are related as conjectured in
Ref.~\cite{duhCascade}.

Another observation is that relaxation in most \diff{homogeneous} microscopic many-body systems
studied in the literature seems to lie in our CMV phase I, where relaxation arises
due to local operators ``escaping to infinity''~\cite{duhCascade}, never
returning. That is, the probability of backflow echo, where an operator
with small support grows to some large support $\ket{n}$ and then shrinks again,
is exponentially suppressed in $n$, as encoded in partial binorms. In this
phase, the decay of correlations is always slower than the decay of Verblunsky
coefficients, $|\lambda|>\mu$, and the truncated propagator gives the correct RP
resonances because we truncate the asymptotically irrelevant information. On the
other hand, in phase II, where the probabilities of backflow echoes are
exponentially increasing in $n$, the eigenvalues of the truncated CMV matrix do
not give the correct RP resonances $\lambda$ (its leading eigenvalue converges
to $\mu \neq \abs{\lambda}$). Due to what seems like
destructive interference with the operator backflow, the decay of correlations is
enhanced, resulting in $|\lambda|<\mu$. It would be interesting to explicitly construct \diff{homogeneous} locally interacting many-body systems lying in phase II. Note that in our coarse-grained interpretation, a basis element
$\ket{n}$ corresponds to operators with support proportional to $n$ sites. Unfolding back to
the physical operator space, where each distinct local operator is an
independent basis element, one could presumably get richer physics than just the
two phases we identify even if Verblunsky coefficients do decay
exponentially on average.

The truncated propagator acts directly on the space of operators, and once one
has a time-evolved operator $A(t)$, it can be used to study different
quantities. The simplest, on which we focused, are two-point correlations
$C(t)=\mel{A}{U^t}{A}=\sum_j c_j \lambda_j^t$, which, besides
$\lambda_j$, contain only coefficients $c_j=\braket{A}{\R_j}\braket{\L_j}{A}$. To
more completely describe quantum relaxation, one could also study higher point
correlation functions, for instance, the ones appearing in out-of-time-order
correlators, and expand them over RP resonances, $O(t) := \ev{A(t) B A(t) B} =$ $
\sum_{i, j} \lambda_i^t \lambda_j^t \braket{\L_i}{A} \braket{\L_j}{A} \ev{\R_i B
\R_j B}$. While a naive guess for the leading decay rate of $O(t) \sim
\lambda_1^{2t}$ might be correct in the generic case, we also see that there are
cross terms between different eigenvectors and that the full algebra of Gamow
vectors could play a bigger role than in $C(t)$. Furthermore, we must keep in
mind that the truncated propagator is a non-normal matrix for which the
diverging expansion coefficients \diff{can} play a major role. 

While the truncated propagator method might be viewed as a heuristic tool to
extract the relaxation times, the rigged Hilbert space approach gives a more
formal framework in which all steps can be rigorously justified. Instead of the
rigged Hilbert space approach one could also use functional analysis to
directly study the properties of infinite-dimensional operators. From a physicist's
standpoint, where one often has to rely on numerics, the truncated propagator
approach offers a way to correctly describe physics in the thermodynamic limit
with a finite-dimensional matrix, without having to explicitly worry about all
the intricacies of infinite-dimensional operators. In other words, it
deals with ``infinities'' in the thermodynamic limit and the non-commutativity of
limits $t \to \infty$ and system size $N \to \infty$. Another framework that
also directly works in the thermodynamic limit is the C$^*$-algebra approach used, e.g.,
in quantum spin chains~\cite{naaijkens2013quantumSpin,Doyon2017,
ampelogiannis2023}. It would be interesting if the formalism presented could be
framed and made useful also in the C$^*$-algebra language.

Our work also reveals several interesting properties and questions within the
mathematics of OPUC. Particularly intriguing are \diff{the effects that the ring of badly conditioned eigenvalues of the truncated CMV matrices has, for instance, hiding the leading RP resonance in phase II as well as the subleading ones in both phases, and} a subtle way in
which one has to treat the limits like that in the \S function, $D(z)=\lim_{n
\to \infty}\frac{1}{\varphi^\star_n(z)}$. Taking a finite order $n$ polynomial
$\varphi^\star_n(z)$, no matter how large $n$ is, is a bad approximation to
$D(z)$ in the sense of giving an incorrect convergence radius and in turn
$\lambda$. Making a ``small perturbation'', like expanding the RHS \diff{or using analytical continuation}, on the other
hand does give the correct $\lambda$. It would be nice to better understand
when, why and how such a situation arises, and if it is in any way related to
pseudospectra, where somewhat similar effects to those in phase II can be observed~\cite{boombust}.

We studied CMV matrices modeling fully chaotic systems without any symmetries
that exhibit exponential decay of all correlation functions of local
operators. Important cases going beyond that are systems with conservation laws
where some power-law decays are expected. While we have recently seen progress
in applying RP resonances and the truncated operator to systems with
conservation laws and connecting them to their transport and hydrodynamic
properties~\cite{3site,prx}, it would be interesting to apply CMV matrices to
systems with symmetries, as well as to any other situation where unitaries in
simple forms are desired, e.g., in various Krylov space based \diff{techniques~\cite{Dimarsky,trunin25,yehPRB26}.}

\section*{Acknowledgements}
UD and MŽ acknowledge Grants No.~N1-0504, No.~J1-70049, and No.~P1-0402 from Slovenian Research Agency (ARIS). FH has received support under the Major Research Program of PSL Research University ``Statistical Physics and Mathematics'' launched by PSL Research University and implemented by ANR with the references ANR-10-IDEX-0001.

\pagebreak
\appendix
\numberwithin{equation}{section}

\section{Exact correlation functions for the shift with an impurity}
\label{app:shift_imp_c}

Correlation functions in shift with an impurity~\eqref{eq:shift_imp} in an
arbitrary state can be computed exactly. For $t \geq 0$ we obtain
\begin{equation}
    \mel{n}{U^t}{m} = \begin{cases}
    \delta_{n, m + 2t} ; &\text{$n, m$ odd} \\
    \delta_{n, m - 2t} ; &\text{$n, m$ even and } n \geq 2 \\
    -\cos \varphi ; &\text{$n$ odd, $m$ even, } m \geq 2 \text{ and } 2t = n + m - 1 \\
    (\sin \varphi)^2 (\cos \varphi)^{t - (n+m+1)/2} ; &\text{$n$ odd, $m$ even, } m \geq 2 \text{ and } 2t \geq n + m + 1 \\
    \sin \varphi (\cos \varphi)^{t - (n+1)/2} ; &\text{$n$ odd, } m = 0 \text{ and } 2t \geq n + 1 \\
    \sin \varphi (\cos \varphi)^{t - m/2}; & n = 0, \text{ $m$ even, } m \geq 2 \text{ and } 2t \geq m \\
    (\cos \varphi)^t ; &m = 0, n = 0 \text{ and } \\
    0 ; &\text{otherwise} \end{cases}
\end{equation}
The result has a nice physical interpretation in both scattering and operator
dynamics picture. The first two lines describe both $n, m$ being stuck in the
tails. The third and fourth line describe the scattering: we get a delta bump
(line 3) followed by an exponential tail (line 4). In the operator dynamics
interpretation, that corresponds to an operator shrinking from large support,
undergoing non-trivial dynamics at small supports and the spreading to large
support again. Lines 5, 6, 7 describe radiation from the $\ket{0}$ state or an
operator at initially small support spreading exponentially.

\section{Analytic structure of the \S function}
\label{app:Dnm}

We are going to analyze the analytic structure of the \S function $D(z)$, focusing on extracting its convergence radius $R_{\rm D}$ (\ref{eq:thm}) from a finite-$n$ polynomials as suggested in Eq.~(\ref{eq:Dnm}).

\subsection{Phase I}
\begin{figure}[t!]
  \centerline{\includegraphics[width=0.99\linewidth]{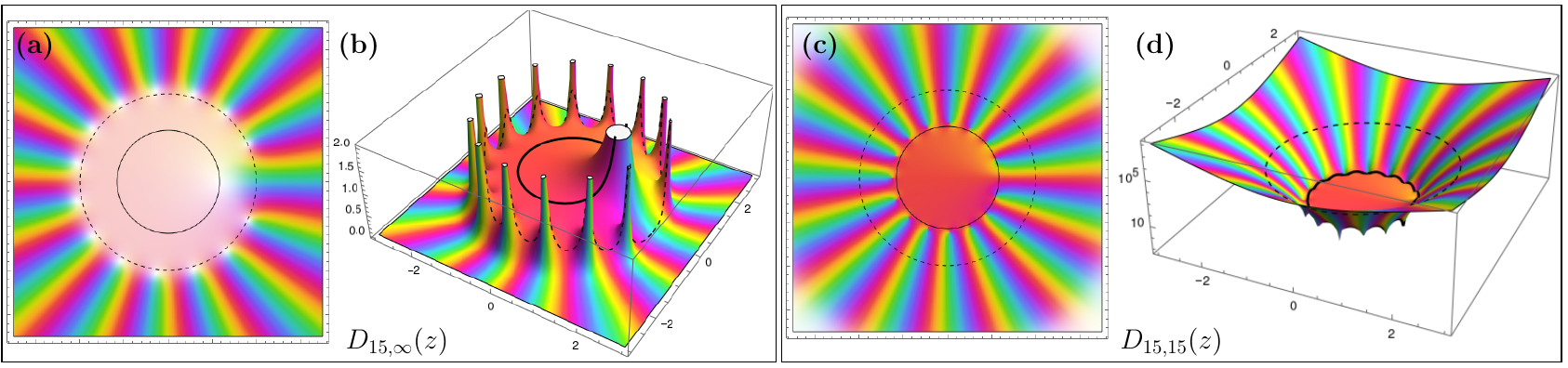}}
  \caption{Phase I at $a=\mu=\frac{1}{2}$. Convergence radius (\ref{eq:thm}) of the \S function $D(z)$ (\ref{eq:Dz}) using finite $n=15$ (\ref{eq:Dnm}). (a) and (b) are for the rational function $\frac{1}{\varphi^\star_{15}(z)}$, whereas (c) and (d) are for a polynomial of order $15$ obtained by Taylor expansion of $\frac{1}{\varphi^\star_{15}(z)}$. Both give the correct $R_D$ (full circle at $R_D=1/\lambda \approx 1.16$). Colors show phases while an additional surface in (b) and (d) shows the norm of $D_{15,\infty}(z)$ and $D_{15,15}(z)$. Dashed circle has radius $1/\mu=2$.}
    \label{fig:DI}
\end{figure}

Let us also illustrate the working of the Theorem in Eq.~(\ref{eq:thm}) that rigorously gives $\lambda$ in terms of the convergence radius $R_D$ of the \S function $D(z)$. In Fig.~\ref{fig:DI}, we show analytic structure in the complex plane $z$ of a finite-$n$ approximation $1/\varphi^\star_n(z)=: D_{n,\infty}$, using small $n=15$ for clarity.
The zeroes of $\varphi^\star_n(z)$ will result in the poles of $D_{n,\infty}(z)$, nicely visible in Figs.~\ref{fig:DI}(a) and (b) (see bottom right Fig.~\ref{fig:presek} for zeroes of $\varphi_n(z)$). As clearly visible in frame (b) the pole nearest to the origin determining $R_{D}$ is at $1/\lambda_1$, and then there is an additional ring of poles at $1/\mu$. An interesting thing though happens when we Taylor expand $D_{15,\infty}$ around $z=0$, keeping only the terms up to for instance order $15$, obtaining $D_{15,15}(z)$ (\ref{eq:Dnm}). Such, in a way small, perturbation results in the ring of poles that were previously at $1/\mu$ (Fig.~\ref{fig:DI}(a) and (b)) disappearing and ``changing'' into a ring of zeroes. Whereas in $D_{n,\infty}(z)$ the convergence radius is determined by a single singularity on the real axis, in $D_{n,n}(z)$ the divergence (for $n \to \infty$) happens at the whole convergence radius (Fig.~\ref{fig:DI}(d)). Note that the appearance of zeroes around the convergence radius is perfectly in line with the Jentzsch theorem~\cite{jentzsch}: if one expands a function $f(z)$ with a finite convergence radius $R$ into a power series to order $n$, the zeroes of the resulting partial sums (finite polynomials) will accumulate uniformly on a circle with radius $R$.

\subsection{Phase II}
\begin{figure}[b!]
  \centerline{\includegraphics[width=0.99\linewidth]{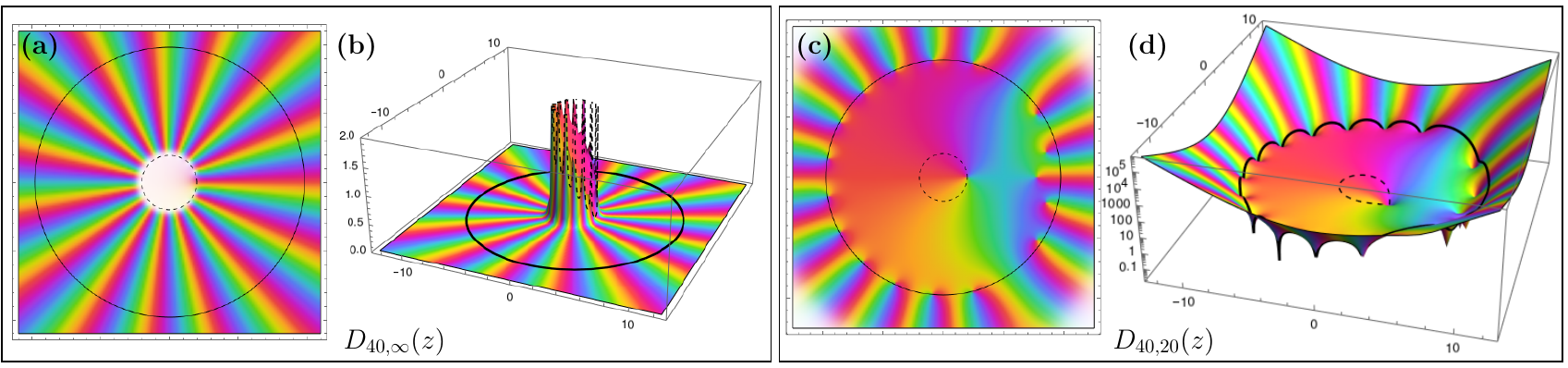}}
  \caption{Phase II at $a=-\frac{2}{5}$ and $\mu=\frac{1}{2}$. Convergence radius (\ref{eq:thm}) of $D(z)$ (\ref{eq:Dz}) using finite $n=40$ (\ref{eq:Dnm}). (a) and (b) are for the rational function $\frac{1}{\varphi^\star_{40}(z)}$, whereas (c) and (d) are for a polynomial of order $20$ obtained by Taylor expansion of $\frac{1}{\varphi^\star_{40}(z)}$. $D_{n,\infty}$ does not give the correct $R_D$ (left), while $D_{40,20}$ does (right). A seemingly innocuous Taylor expansion radically changes a finite-$n$ approximation of the \S function: (a) and (b) vs. (c) and (d). Colors show phases of complex functions while an additional surface in (b) and (d) shows the norm. Full circle is at the expected correct convergence radius $R_D=1/\lambda \approx 9.75$, while the dashed one has radius $1/\mu=2$.}
    \label{fig:DII}
\end{figure}

Let us look at the \S function and a finite-$n$ approximation obtained by taking $U_{N=40}$, calculating $\Phi_{40}(z)=\det{(z-U_{40})}$, and then getting the reversed polynomial $\Phi^\star_{40}(z)$ and $D_{40,\infty}(z):=1/\varphi^\star_{40}(z)$. Results are shown in Fig.~\ref{fig:DII} in frames (a) and (b).
We can see that in-line with all zeroes of $\Phi_{40}(z)$ being on a circle with radius $\mu=1/2$ (like in the bottom middle Fig.~\ref{fig:presek}), $D_{40,\infty}(z)$ has a ring of poles at radius $2$, which, however, is not at the correct $R_D$ (full circle). Expanding $D_{40,\infty}(z)$ to order $m=20$, obtaining $D_{40,20}(z)$ (\ref{eq:Dnm}), we can see in frames (c) and (d) that now the divergence (in $n,m \to \infty$) happens at the correct radius. Similarly as in Fig.~\ref{fig:DI}, the ring of poles at $1/\mu$ changes into a ring of zeroes at $R_D$, correctly indicating the convergence radius (Jentzsch theorem~\cite{jentzsch}). This high sensitivity of the poles and zeroes of $D_{n,\infty}(z)$ and $\Phi_n(z)$, respectively, to small perturbations is a reflection of a badly conditioned truncated matrix $U_N$. A condition number $\kappa(A)$ of a matrix $A$ tells us how sensitive is $A$ to small perturbations and is for diagonalizable $A$ equal to $\kappa(A)=\nor{V}\,\nor{V^{-1}}$, where $V$ diagonalizes $A$, $A=VDV^{-1}$~\cite{trefethen}. The condition number $\kappa$ for instance bounds the eigenvalue change $|\Delta \lambda_j| \le \kappa(A) \nor{\delta A}$ under perturbation $\delta A$. Taking a 2-norm one has $\kappa(A)=s_{\rm max}(V)/s_{\rm min}(V)$ where $s$ are the singular values. For our truncated $U_N$ the singular values of $V$ are complicated, however we can get an estimate using singular values of $U_N$ itself. Namely, $U_N$ is a rank 1 perturbation\footnote{$U_N$ and unitary CMV of size $N$ differ only in the last column.} of a finite unitary CMV matrix of size $N$, and therefore all but one singular values of $U_N$ are $1$, with the single nontrivial one being $s_{\rm min}(U_N)=2a\mu^N$. This exponentially small singular value causes high sensitivity of $U_N$. Nevertheless, it seems to play a role only in phase II.

\section{Transfer matrix calculations}
\label{app:cmv_tm}

\subsection{Solution of the inhomogeneous eigenequation}
\label{app:cmv_ev}

We are solving
\begin{equation}
    U\ket{u} = z \ket{u} - \ket{\phi},
\end{equation}
where $U = \cmvL \cmvM$ is a CMV matrix.
Defining an auxiliary degree of freedom
\begin{equation}
    \ket{v} := \cmvM \ket{u},\label{equ:resolvent_vdef}
\end{equation}
we can write
\begin{equation}
    \cmvL\ket{v} = z\ket{u} - \ket{\phi}.\label{equ:resolvent_vu}
\end{equation}

Expressing \eqref{equ:resolvent_vdef} and \eqref{equ:resolvent_vu} in components we obtain for $n\geq 1$
\begin{align}
    v_0 &= u_0\\
    v_{2n-1} &= \alpha_{2n-1}^* u_{2n-1} + \rho_{2n-1} u_{2n}\\
    v_{2n} &= \rho_{2n-1}u_{2n-1} - \alpha_{2n-1}u_{2n}
\end{align}
and for $n\geq 0$
\begin{align}
    \alpha_{2n}^* v_{2n} + \rho_{2n}v_{2n+1} &= zu_{2n}\phi_{2n}\\
    \rho_{2n} v_{2n} -\alpha_{2n}v_{2n+1} &= zu_{2n+1} - \phi_{2n+1}.
\end{align}

We can compactly write this as
\begin{align}
    v_0 &= u_0\\
    \begin{pmatrix}
        u_{2n}\\
        v_{2n}
    \end{pmatrix} &= \T_{2n-1}(z) \begin{pmatrix}
        u_{2n-1}\\
        v_{2n-1}
    \end{pmatrix}\\
    \begin{pmatrix}
        u_{2n+1}\\
        v_{2n+1}
    \end{pmatrix} &= \T_{2n}(z) \begin{pmatrix}
        u_{2n}\\
        v_{2n}
    \end{pmatrix} + \begin{pmatrix}
        \tfrac{\alpha_{2n}}{z\rho_{2n}} \phi_{2n} + \tfrac{1}{z}\phi_{2n+1}\\
        -\tfrac{1}{\rho_{2n}}\phi_{2n}
    \end{pmatrix},
\end{align}
where the transfer matrices $\T_k(z)$ are defined in Eq.~\eqref{eq:cmv_tm}. By
using the product of transfer matrices defined in Eq.~\eqref{eq:cmv_mm}, we can
explicitly solve this as
\begin{align}
    \begin{pmatrix}
        u_{2n} \\ v_{2n}
    \end{pmatrix} &= \M_{2n}(z) \vep v_0  + \sum_{k = 0}^{n-1} \M_{2n}(z) \M^{-1}_{2k+1}(z) \begin{pmatrix}
        \tfrac{\alpha_{2k}}{z\rho_{2k}} \phi_{2k} + \tfrac{1}{z}\phi_{2k+1}\\
        -\tfrac{1}{\rho_{2k}}\phi_{2k}
    \end{pmatrix}\\
    &= \M_{2n}(z) \vep v_0 + \M_{2n}(z) \sum_{k = 0}^{n-1}  \M^{-1}_{2k+1}(z) \Big[-\tfrac{1}{z}\T_{2k}(z)\ve_1 \phi_{2k} + \tfrac{1}{z}\ve_1\phi_{2k+1}\Big]\\
    &= \M_{2n}(z) \vep v_0 - \frac{1}{z}\M_{2n}(z) \sum_{k = 0}^{2n-1} (-1)^k \M^{-1}_{k}(z) \ve_1 \phi_{k}\\
     \begin{pmatrix}
        u_{2n+1} \\ v_{2n+1}
    \end{pmatrix} &= \M_{2n+1}(z) \vep v_0 - \frac{1}{z}\M_{2n+1}(z) \sum_{k = 0}^{2n-1} (-1)^k \M^{-1}_{k}(z) \ve_1 \phi_{k}
\end{align}
which we can compactly write as
\begin{equation}
    \begin{pmatrix}
        u_{n} \\ v_{n}
    \end{pmatrix} = \M_{n}(z) \vep v_0 - \frac{1}{z}\M_{n}(z) \sum_{k = 0}^{2\lfloor n/2\rfloor-1} (-1)^k \M^{-1}_{k}(z) \ve_1 \phi_{k}.
\end{equation}
We still have one free parameter $v_0$, which can be fixed by an appropriate boundary condition.

\subsection{Orthogonal polynomials expressed with transfer matrices}
\label{app:cmv_opuc}

We have the following relationships between OPUC and transfer matrices
\begin{align}
     \M_n(z) \vep &= \begin{pmatrix}
        \xopuc_n(z) \\
        \chi_n(z)
    \end{pmatrix},  \label{eq:cmv_tm_opuc_m1} \\
     (-1)^n \vem \M_n^{-1}(z) &= \begin{pmatrix}
        \chi_n(z) \\
        -\xopuc_n(z)
    \end{pmatrix}, \label{eq:cmv_tm_opuc_m2} \\
    \M_k(z) \vem &= \begin{cases}
        z^{-k/2} \begin{pmatrix}
            \psi_k(z) \\
            -\psi_k^\star(z)
        \end{pmatrix}; &k \text{ even} \\
        \begin{pmatrix}
            -z^{-(k + 1)/2} \psi^\star_k(z) \\
            z^{-(k - 1)/2} \psi_k(z)
        \end{pmatrix}; &k \text{ odd}
    \end{cases} \label{eq:cmv_tm_opuc_m3}
\end{align}
where $\chi$ and $\xopuc$ are defined in Eqs.~\eqref{eq:chi_def}, \eqref{eq:x_def} and $\psi$ in Eq.~\eqref{eq:psi_def}.
Eqs.~\eqref{eq:cmv_tm_opuc_m1}, \eqref{eq:cmv_tm_opuc_m2}, \eqref{eq:cmv_tm_opuc_m3} can be proven by induction.

% \begin{align}
%     \frac{1}{z^n}\begin{pmatrix}
%         \varphi_{2n}(z) \\
%         \varphi_{2n}^\star(z)
%     \end{pmatrix} &= \M_{2n}(z) \vep, \label{eq:cmv_tm_opuc_1} \\
%     \begin{pmatrix}
%         \frac{1}{z^{n + 1}} \varphi_{2n + 1}^\star(z) \\
%         \frac{1}{z^{n}}\varphi_{2n + 1}(z)
%     \end{pmatrix} &= \M_{2n + 1}(z) \vep. \label{eq:cmv_tm_opuc_2}
% \end{align}
% \begin{align}
%     \frac{1}{z^n}\begin{pmatrix}
%         \varphi_{2n}^\star(z) \\
%         -\varphi_{2n}(z)
%     \end{pmatrix} &= \vem \M_{2n}(z), \label{eq:cmv_tm_opuc_1} \\
%     \begin{pmatrix}
%         -\frac{1}{z^{n}} \varphi_{2n + 1}(z) \\
%         \frac{1}{z^{n + 1}}\varphi_{2n + 1}^\star(z)
%     \end{pmatrix} &= \vem \M_{2n + 1}(z). \label{eq:cmv_tm_opuc_2}
% \end{align}

Let us prove Eq.~\eqref{eq:cmv_tm_opuc_m1}. Writing out the equation
with normalized OPUC $\varphi_k(z)$,
\begin{align}
    \frac{1}{z^n}\begin{pmatrix}
        \varphi_{2n}(z) \\
        \varphi_{2n}^\star(z)
    \end{pmatrix} &= \M_{2n}(z) \vep, \label{eq:cmv_tm_opuc_1} \\
    \begin{pmatrix}
        \frac{1}{z^{n + 1}} \varphi_{2n + 1}^\star(z) \\
        \frac{1}{z^{n}}\varphi_{2n + 1}(z)
    \end{pmatrix} &= \M_{2n + 1}(z) \vep. \label{eq:cmv_tm_opuc_2}
\end{align}
The base case \underline{$N = 0$} is valid by definition of $\varphi_0(z) = 1$
(see Sec.~\ref{sec:opuc}),
\begin{equation}
    \M_0(z) \vep = \begin{pmatrix}
        1 \\ 1
    \end{pmatrix} = \begin{pmatrix}
        \varphi_0(z) \\
        \varphi_0^\star(z)
    \end{pmatrix}.
\end{equation}
We now assume that Eqs.~\eqref{eq:cmv_tm_opuc_1}, \eqref{eq:cmv_tm_opuc_2} are valid
up to $N - 1$ and prove that this implies their validity for $N$. We first consider
even \underline{$N = 2n$},
\begin{align}
    \M_{2n}(z) \vep &= \T_{2n - 1}(z) \begin{pmatrix}
        \frac{1}{z^{n}} \varphi_{2n - 1}^\star(z) \\
        \frac{1}{z^{n - 1}}\varphi_{2n - 1}(z)
    \end{pmatrix} = \frac{1}{z^n \rho_{2n - 1}} \begin{pmatrix}
         - \alpha_{2n - 1}^* \varphi_{2n - 1}^\star(z) + z \varphi_{2n - 1}(z)\\
        - \alpha_{2n - 1} z \varphi_{2n - 1}(z) + \varphi_{2n - 1}^\star(z) 
    \end{pmatrix} \nonumber \\ &= \frac{1}{z^n}\begin{pmatrix}
        \varphi_{2n}(z) \\
        \varphi_{2n}^\star(z)
    \end{pmatrix},
\end{align}
where we used Eq.~\eqref{eq:opuc_iter} in the last line. A similar calculation
can be done for odd \underline{$N = 2n + 1$},
\begin{align}
    \M_{2n + 1}(z) \vep &= \T_{2n}(z) \frac{1}{z^n}\begin{pmatrix}
        \varphi_{2n}(z) \\
        \varphi_{2n}^\star(z)
    \end{pmatrix} = \frac{1}{\rho_{2n}} \begin{pmatrix}
        \frac{1}{z^{n + 1}} \left[  - \alpha_{2n} z \varphi_{2n}(z) + \varphi^\star_{2n}(z)\right] \\
        \frac{1}{z^n} \left[ z \varphi_{2n}(z)-\alpha_{2n}^* \varphi_{2n}^\star(z)\right]
    \end{pmatrix} \\
    &= \begin{pmatrix}
        \frac{1}{z^{n+1}}\varphi_{2n + 1}^\star(z) \\
        \frac{1}{z^n} \varphi_{2n + 1}(z)
    \end{pmatrix},
\end{align}
where we again used Eq.~\eqref{eq:opuc_iter} in the last line. This concludes
the proof.

Eqs.~\eqref{eq:cmv_tm_opuc_m2}, \eqref{eq:cmv_tm_opuc_m3} can be proven in a similar fashion.

\subsection{Further details for the derivation of the resolvent}
\label{app:cmv_resolvent}

We are continuing the derivation in Sec.~\ref{sec:cmv_resolvent} from
the boundary condition we deduced from the asymptotic behavior of transfer
matrices in Eq.~\eqref{eq:t_asymp}. Namely, the appropriate boundary
condition at infinity assuming $\abs{z} > 1$ is
\begin{equation}
    u_{2N} =  \ve_1\M_{2N}(z) \left[\vep v_0 - \frac{1}{z} \sum_{k = 0}^{2N-1} (-1)^k \M^{-1}_{k}(z) \ve_1 \phi_{k}\right] \to 0.
\end{equation}
In in order to take the $N\to \infty$ limit of this expression, let us first write
\begin{align}
    \tilde{\M}_{2N}(z) &= \tfrac{1}{z^N} \M_{2N}(z).
\end{align}
and then define
\begin{align}
    \tilde{\M}_{\infty}(z) &= \lim_{N\to \infty}\tilde{\M}_{2N}(z).
\end{align}
As opposed to $\lim_{N\to \infty} \M_{2N}(z)$, $\tilde{\M}_{\infty}(z)$ is indeed a finite matrix.
Thus, we can write
\begin{align}
    \ve_1\tilde{\M}_{\infty}(z) \left[\vep v_0 - \frac{1}{z} \sum_{k = 0}^{\infty} (-1)^k \M^{-1}_{k}(z) \ve_1 \phi_{k}\right] &= 0
\end{align}
from which we can identify $v_0$ as
\begin{align}
    v_0 &= \frac{1}{z \ve   _1\tilde{\M}_{\infty}(z) \vep}\left[\sum_{k = 0}^{\infty} (-1)^k \ve   _1\tilde{\M}_{\infty}(z) \M^{-1}_{k}(z) \ve_1 \phi_{k}\right]. \label{eq:cmv_v0}
\end{align}
Now using the boundary condition \eqref{eq:cmv_v0} in Eq.~\eqref{eq:cmv_iter_res}
\begin{align}
    \begin{pmatrix}
        u_{n} \\ v_{n}
    \end{pmatrix} &= \frac{1}{z}\M_{n}(z)\vep \frac{1}{\ve   _1\tilde{\M}_{\infty}(z) \vep} \left[\sum_{k = 0}^{\infty} (-1)^k \ve   _1\tilde{\M}_{\infty}(z)\1 \M^{-1}_{k}(z) \ve_1 \phi_{k}\right]\nonumber\\
    &\quad - \frac{1}{z}\M_{n}(z) \sum_{k = 0}^{2\lfloor n/2\rfloor-1} (-1)^k \M^{-1}_{k}(z) \ve_1 \phi_{k}\\
    &= \frac{1}{2z}\M_{n}(z)\vep \frac{\ve   _1\tilde{\M}_{\infty}(z)\vem}{\ve_1\tilde{\M}_{\infty}(z) \vep} \left[\sum_{k = 0}^{\infty} (-1)^k \vem\M^{-1}_{k}(z) \ve_1 \phi_{k}\right]\nonumber\\
    &\quad+\frac{1}{2z}\M_{n}(z)\vep\left[\sum_{k = 0}^{\infty} (-1)^k \vep\M^{-1}_{k}(z) \ve_1 \phi_{k}\right] - \frac{1}{z}\M_{n}(z) \sum_{k = 0}^{2\lfloor n/2\rfloor-1} (-1)^k \M^{-1}_{k}(z) \ve_1 \phi_{k}
\end{align}
where we inserted $\1 = \frac{1}{2} (\vep \vep^\top + \vem \vem^\top)$ in the
first line. We can now first evaluate the matrix element of the resolvent
\begin{align}
    &\mel{n}{R(z)}{m} = \braket{n}{u\vert_{\phi_k = \delta_m, k}}  \\
    &= \frac{1}{2z}\ve_1\M_{n}(z)\vep \frac{\ve   _1\tilde{\M}_{\infty}(z)\vem}{\ve_1\tilde{\M}_{\infty}(z) \vep} (-1)^m \vem\M^{-1}_{m}(z) \ve_1\nonumber\\
    &\quad+\frac{1}{2z}\ve_1\M_{n}(z)\vep(-1)^m \vep\M^{-1}_{m}(z) \ve_1 - \frac{1}{z}\ve_1\M_{n}(z) \theta(m\leq 2\lfloor n/2\rfloor-1)(-1)^m \M^{-1}_{m}(z) \ve_1
\end{align}
%where $\vartheta(x) = \begin{cases}
%    1;& x > 0 \\
%    0;& \text{otherwise}
%\end{cases}$ is the step function.
where $\vartheta(x\leq y) = \begin{cases}
    1;& x\leq y \\
    0;& \text{otherwise}
\end{cases}$ is a step function.
This leads to the complete result
\begin{align}
    R(z) &= \left(\sum_{n = 0}^\infty \ve_1 \M_{n}(z) \vep \ket{n}\right) \left(\frac{1}{2z}\frac{\ve_1 \tilde{\M}_{\infty}(z)\vem}{\ve_1 \tilde{\M}_\infty(z)\vep}\right) \left( \sum_{m = 0}^\infty (-1)^m \vem\M^{-1}_m(z) \ve_1 \bra{m} \right) + \Gamma(z) \\
    \Gamma(z) &:= \sum_{n,m=0}^\infty\Big[\frac{1}{2z}\ve_1\M_{n}(z)\vep(-1)^m \vep\M^{-1}_{m}(z) \ve_1\label{eq:cmv_zeta}\\
    &\hspace{90pt}- \frac{1}{z}\ve_1\M_{n}(z) \theta(m\leq 2\lfloor n/2\rfloor-1)(-1)^m \M^{-1}_{m}(z) \ve_1\Big] \ket{n}\bra{m}\nonumber
\end{align}
Recall that our derivation was valid for $\abs{z} > 1$. If we instead consider $\abs{z} < 1$,
we must only change the ``boundary condition at infinity'', i.e., according to
Eq.~\eqref{eq:t_asymp}, we must now fine-tune $v_{2N \to \infty}$ to vanish
instead. Redoing the same derivation in that case, leads to the complete result
written in Eq.~\eqref{eq:cmv_resolvent}, where we have additionally pulled the
limit $N \to \infty$ in front of the fraction.

\subsection{Asymptotic expansion of transfer matrices}
\label{app:cmv_asymp}

We are pursuing an asymptotic expression for $\M_{n \to \infty}(z)$ in the case
of exponentially decaying $\alpha_k = a \mu^k$. The idea is to treat all
$\alpha_k$ to be of comparable order, which is valid only for $k \gg 1$, since
$\alpha_{k_1} \gg \alpha_{k_2}^2$ for any $k_1 \sim k_2$, $k_1 \gg 1$, $k_2 \gg
1$. This will give us an asymptotic expression for $\M_{n \to \infty}(z)$ acting
on some initial vector that will include constant set by $\alpha_k$ for small
$k$ where our approximation is not valid yet.

Let us first expand the product of two successive transfer matrices. It is easy
to see that to first order in $\alpha_k$
\begin{equation}
    \T_{2n-1}(z) \T_{2n - 2}(z) = \underbrace{\begin{pmatrix}
        z & 0 \\ 0 & 1/z
    \end{pmatrix}}_{:= \tilde \T_b} + \underbrace{\begin{pmatrix}
        0 & - \alpha_{2n - 2}^* - \alpha_{2n-1}^*/z \\ -z \alpha_{2n-1} - \alpha_{2n - 2} & 0
    \end{pmatrix}}_{:= \tilde \T_{2n-1}} + \mathcal{O}(\alpha_k^2).
\end{equation}
Using this, we can obtain the first order expression for the product of multiple
consecutive transfer matrices
\begin{align}
    \M_{2N}(z) \M^{-1}_{2N'}(z) &= \mathop{\overleftarrow{\prod}}\limits_{k = 2N'+1}^{2N} \T_{k-1}(z)  = \tilde \T_b^{N - N'} + \sum_{n = N' + 1}^N \tilde \T_b^{N - n} \tilde \T_{2n-1} \tilde \T_b^{n - N' - 1} + \mathcal{O}(\alpha_k^2) \nonumber \\
    &= \begin{pmatrix}
        z^{N - N'} & -z^{N + N'} \sum_{k = 2N' + 1}^{2N}  \tfrac{\alpha_{k-1}^*}{z^k} \\
        - \frac{1}{z^{N + N'}} \sum_{k = 2 N' + 1}^{2N} \alpha_{k-1} z^k & \frac{1}{z^{N - N'}}
    \end{pmatrix} + \mathcal{O}(\alpha_k^2) \label{eq:m_asmyp_even}
\end{align}
and similarly for odd indices
\begin{equation}
    \M_{2N+1}(z) \M^{-1}_{2N'}(z) = \begin{pmatrix}
        - z^{N + N' + 1} \sum_{k = 2 N' + 1}^{2N + 1} \frac{\alpha_{k-1}^*}{z^k} & z^{N - N' + 1} \\
        \frac{1}{z^{N - N' + 1}} & -\frac{1}{z^{N + N' + 1}} \sum_{k = 2N' + 1}^{2N + 1} \alpha_{k-1} z^k 
    \end{pmatrix} + \mathcal{O}(\alpha_k^2), \label{eq:m_asmyp_odd}
\end{equation}
both valid for $N \geq N' \gg 1$.

We are interested in $\M_{n \to \infty}(z) \ve_\pm$. We can use
Eqs.~\eqref{eq:m_asmyp_even}, \eqref{eq:m_asmyp_odd} from some large $2N'$
onwards and package the contributions from small indices into some $z$ dependent
constants. Namely, we write
\begin{equation}
    \M_{2N'}(z) \ve_\pm =: \begin{pmatrix}
        \tilde c_1^{(\pm)} \\
        \tilde c_2^{(\pm)}
    \end{pmatrix}
\end{equation}
and then
\begin{align}
    \M_{2N \to \infty}(z) \ve_\pm &= \M_{2N \to \infty}(z)\M^{-1}_{2N'}(z)\M_{2N'}(z) \ve_\pm = \nonumber \\
    &=\begin{pmatrix}
        \tilde c_1^{(\pm)}z^{N - N'} - \tilde c_2^{(\pm)}z^{N + N'} \sum_{k = 2N' + 1}^{2N} \alpha_{k-1}^* \tfrac{1}{z^k} \\
        -\tilde c_1^{(\pm)} \frac{1}{z^{N + N'}} \sum_{k = 2 N' + 1}^{2N} \alpha_{k-1} z^k + \tilde c_2^{(\pm)} \frac{1}{z^{N - N'}}
    \end{pmatrix} + \mathcal{O}(\alpha_k^2) \nonumber \\
    &= \begin{pmatrix}
        z^N \left[c_1^{(\pm)} - c_2^{(\pm)} \sum_{k = 0}^{2N-1} \alpha_{k}^* \tfrac{1}{z^k}\right] \\[0.5em]
       \frac{1}{z^N} \left[-c_1^{(\pm)} \sum_{k = 0}^{2N-1} \alpha_{k}z^k + c_2^{(\pm)}\right]
    \end{pmatrix} + \mathcal{O}(\alpha_k^2), \label{eq:cmv_asymp_even}
\end{align}
where we have pulled all the $N'$ dependent constant into $c_{1, 2}^{(\pm)}$.
Explicitly, $c_{1, 2}^{(\pm)}$ now depends on $N', z$ and $\alpha_{1 \leq k \leq
N'}$, which are all finite numbers not affecting the asymptotics. Similarly,
the result for odd indices is
\begin{equation}
    \M_{2N + 1 \to \infty}(z) \ve_\pm = \begin{pmatrix}
        \frac{1}{z^{N + 1}} \left[-c_1^{(\pm)} \sum_{k = 0}^{2N} \alpha_{k} z^k + c_2^{(\pm)}\right] \\[0.5em]
        z^{N + 1} \left[c_1^{(\pm)} - c_2^{(\pm)} \sum_{k = 0}^{2N} \alpha_{k}^* \tfrac{1}{z^k} \right]
    \end{pmatrix} + \mathcal{O}(\alpha_k^2). \label{eq:cmv_asymp_odd}
\end{equation}

\pagebreak
\bibliography{references}

\end{document}